\documentclass{aa}

\usepackage{graphicx}
\usepackage{txfonts}

\newcommand{\be}{\begin{equation}}
\newcommand{\ee}{\end{equation}}
\newcommand{\ba}{\begin{eqnarray}}
\newcommand{\ea}{\end{eqnarray}}
\newcommand{\bc}{\begin{center}}
\newcommand{\ec}{\end{center}}

\begin{document}

\title{Hadronic processes within collective stellar winds}

\author{Eva Domingo-Santamar\'{\i}a$^{1}$ \& Diego F. Torres$^{2,3}$  }

\institute{Institut de F\'{\i}sica d'Altes Energies (IFAE),
Edifici C-n, Campus UAB, 08193 Bellaterra, Spain.
\and Institut de Ciencies de l'Espai (ICE),
  Campus UAB,
  Facultat de Ciencies, Torre C5 - parell - 2a planta
  08193 Barcelona, Spain
  \and Lawrence Livermore National Laboratory, 7000 East Ave.,
Livermore 94550, USA }

\authorrunning {Domingo-Santamar\'{\i}a \& Torres}

\titlerunning{Hadronic processes in winds}

\abstract{Recently, we have proposed that the interaction between
relativistic protons resulting from Fermi first order acceleration
in the superbubble of a stellar OB association or in other nearby
accelerator and ions residing in single stellar winds of massive
stars could lead to TeV sources without strong counterparts at
lower energies. Here we refine this analysis in several
directions. We study collective wind configurations produced by a
number of massive stars, and obtain densities and expansion
velocities of the stellar wind gas that is to be target of
hadronic interactions. We study the expected $\gamma$-ray emission
from these regions, considering in an approximate way the effect
of cosmic ray modulation. We compute secondary particle production
(electrons from knock-on interactions and electrons and positrons
from charged pion decay), and solve the loss equation with
ionization, synchrotron, bremsstrahlung, inverse Compton, and
expansion losses. We provide examples where configurations can
produce sources for GLAST satellite, and the MAGIC/HESS/VERITAS
telescopes in non-uniform ways, i.e., with or without the
corresponding counterparts. We show that in all cases we studied
no EGRET source is expected. Finally, we comment on HESS J1303-631
and on Cygnus OB 2 and Westerlund 1 as two associations where this
scenario could be tested.}

\maketitle

\section{Introduction}

Early-type stellar associations have long been proposed as cosmic
rays  acceleration sites (Cesarsky \&  Montmerle 1983,  Manchanda
et al. 1996, Bozhokin \& Bykov 1994,  Parizot et al. 2004, also
Dorman 1999, Romero et al. 1999). For instance, it is expected
that collective effects of strong stellar winds and supernova
explosions at the core of the associations will produce a
large-scale shock (the supperbubble region) which will accelerate
particles up to energies of hundreds of TeV (e.g., Bykov \&
Fleishman 1992a,b; Bykov 2001). Such relativistic particles, if
colliding with a dense medium, may produce significant
$\gamma$-ray emission, mainly through hadronic interactions.

O, B, and WR stars, lose a significant fraction of their masses in
their winds. Indeed, the ultimate result of a stellar wind with a
high mass-loss rate is to give back gas mass to the interstellar
medium (ISM). For the sake of this discussion, we make here a
distinction  between the mass that is yet contained in single or
collective winds of massive stars (in movement) and the mass that
is free in the ISM (at rest). If star formation is on-going, the
latter would greatly dominate, since only a fraction of the total
gas mass contained in the association is to end in stars. However,
when the star formation is coeval and is currently ended, and
particularly if one or several supernova explosions have pushed
away the free gas mass of the region, or when the stars under
consideration are located in the outskirts of a larger
association, the mass contained in the innermost regions of the
winds can exceed that contained in a similarly sized area of the
ISM. When computing hadronic $\gamma$-ray luminosities, the mass
in winds can not be considered negligible in these situations.

Consequently, the winds of a group of massive stars, particularly
if located close to a cosmic ray acceleration region, may act as
an appropriate cosmic ray target (e.g., Romero \& Torres 2003).
And, because of the wind modulation of the incoming cosmic ray
flux, what we discuss in more detail below, only high energy
particles will be able to penetrate into the wind. Thus, only high
energy photons might be generated copiously enough as to be
detected. The proposed scenario may then predict new potential
sources for the new generation of ground-based \v{C}erenkov
telescopes, which would at the beginning be unidentified due of
their lack of lower energy counterpart.\footnote{For an
alternative scenario for producing sources in the GeV-TeV regime
in non-uniform ways see Bosch-Ramon et al. (2005).} In view of the
several unidentified sources already detected by HESS in the
hundred of GeV -- TeV energy regime (Aharonian et al. 2005), the
aim of this work is to further evaluate this possibility.


\section{The gas within a collective wind}

We adopt a similar modelling to that of Cant\'o et al. (2000) (see
also Chevalier \& Clegg 1985, Ozernoy et al. 1997, Stevens  \&
Hartwell 2003) to describe the wind of a cluster (or a
sub-cluster) of stars. { This is a hydrodynamical model that does
considers the effects of magnetic fields, see next section for
further discussion on the expected (low) values of the magnetic
field in the collective wind region.} Consider that there are $N$
stars in close proximity, uniformly distributed within a radius
$R_c$, with a number density
\be \label{n} n = \frac{3N}{ 4\pi R_c^3} \,. \ee
Each star has its own mass-loss rate ($\dot M_{i}$) and (terminal)
wind velocity ($V_i$), and average values can be defined as
\ba \dot M_w &=& \frac{1}{N} \sum_{i}^{N} \dot M_i \,,  \\
V_w &=& \left( \frac{\sum_{i}^{N} \dot M_i {V_i}^2}{N\,\dot M_w}
\right)^{1/2} \,. \ea
All stellar winds are assumed to mix with the surrounding ISM and
with each other, filling up the intra-cluster volume with a hot,
shocked, collective stellar wind. A stationary flow in which mass
and energy provided by stellar winds escape through the outer
boundary of the cluster is established.
For an arbitrary distance $R$ from the center of the cluster, mass
and energy conservation imply that
\ba \frac{4 \pi}{3} R^3 n \dot M_w &=& 4 \pi R^2 \rho V \,, \\
\frac{4 \pi}{3} R^3 n \dot M_w \left( \frac 12 {V_w}^2 \right)&=&
4 \pi R^2 \rho V \left( \frac 12 {V}^2 + h\right) \,, \ea
where $\rho$ and $V$ are the mean density and velocity of the cluster
wind flow at position $R$ and $h$ is its specific enthalpy (sum of
internal energy plus the pressure times the volume),
\be h = \frac{\gamma}{\gamma-1} \frac{P}{\rho} \,, \ee
with $P$ being the mean pressure of the wind and $\gamma$ being
the adiabatic index (hereafter $\gamma=5/3$ to fix numerical
values). From the mass conservation equation we obtain,
\be \label{mm} \rho V = \frac {n \dot M_w}{3}R \,, \ee
whereas the ratio of the two conservation equations imply
\be \label{fff} \frac 12 V^2 + h = \frac 12 {V_w}^2 \,. \ee
The equation of motion of the flow is
\be \label{motion} \rho V \frac {dV}{dR} =
-\frac{dP}{dR} - n \dot M_w V \,, \ee
which, introducing the adiabatic sound speed $c$,
\be \label{cc} c^2 = \gamma \frac {P}{\rho} \,, \ee
can be written as
\be \label{ddd} \rho V \frac {dV}{dR} = - \frac 1\gamma
\frac{d(\rho c^2)}{dR} - n \dot M_w V \,. \ee
From the definition of enthalpy and Equation (\ref{fff}),
the adiabatic sound speed can be expressed as
\be \label{ccc} c^2 = \frac {\gamma-1}{2} ({V_w}^2-V^2) \,. \ee
Using Equation (\ref{mm}), its derivative $d\rho$ and Equation
(\ref{ccc}) in (\ref{ddd}) one obtains
\be \frac {dR}{R} = \frac {dV}{V} \left[
\frac{(\gamma-1){V_w}^2-(\gamma+1)V^2}
{(\gamma-1) {V_w}^2 + (5\gamma+1)V^2}\right] \,, \ee
which can be integrated and expressed in more convenient
dimensionless variables ($v\equiv V/V_w$ and $r \equiv R/R_c$) as
follows
\be v \left[ 1+\frac {5\gamma+1}{\gamma-1}
v^2\right]^{-(3\gamma+1)/(5\gamma+1)} = Ar \,, \label{inside} \ee
with $A$ an integration constant.\\
When $R>R_c$, i.e., outside the cluster, by definition $n$ is
equal to 0, and the mass conservation equation is
\be \dot M_{\rm assoc} \equiv \frac{4\pi}{3} {R_c}^3 n \dot M_w =
4\pi R^2 \rho V \,, \label{eee} \ee
where the middle equality gives account of the contribution of all
stars in the association, and $\dot M_{\rm assoc} = \sum_i \dot
{M_i}$ is the mass-loss
rate at the outer boundary $R_c$.\\
Substituting Equation (\ref{ccc}) and (\ref{eee}) into the $n=0$
realization of Equation (\ref{ddd}) one obtains
\be -\frac {dR}{R} = \frac {dV}{V} \left[
\frac{(\gamma-1){V_w}^2-(\gamma+1)V^2} {2 (\gamma-1) (V_w^2 -
V^2)}\right] \,, \ee
and integrating, the velocity in this outside region is implicitly
defined from
\be v (1-v^2)^{1/(\gamma-1)} = B r^{-2}, \label{outside} \ee
with $B$ an integration constant.\\
Having constants $A$ and $B$ in Equations (\ref{inside}) and
(\ref{outside}), see below, the velocity at any distance from the
association center can be determined by numerically solving its
implicit definitions, and hence the density is also determined,
through Equation (\ref{mm}) or (\ref{eee}).

From Equation (\ref{outside}), two asymptotic branches can be
found. When $r \rightarrow \infty$, either $v  \rightarrow 0 $
(asymptotically subsonic flow) or $v  \rightarrow 1 $
(asymptotically supersonic flow) are possible solutions. The first
one (subsonic) produces the following limits for the density, the
sound speed and the pressure
\ba
\rho_\infty &=& \frac{\dot M_{\rm assoc}}{4\pi B {R_c}^2 V_w} \,, \\
c_\infty^2 &=& \frac{\gamma-1}{2} {V_w}^2 \,, \\
P_\infty &=& \frac{\gamma-1}{2 \gamma} \frac{\dot M_{\rm assoc}V_w}
{4\pi B {R_c}^2 } \,. \ea
%
%
%
From the latter Equation, if $P_\infty$ is the ISM
pressure far from the association, the constant $B$ can be
obtained as
\be \label{pin} B = \frac{\gamma-1}{2 \gamma}
\frac{\dot M_{\rm assoc}V_w}{4\pi P_\infty {R_c}^2 } \,. \ee

The velocity of the flow at the outer radius $r=1$ follows from
Equation (\ref{outside})
\be v_{r=1} (1-{v_{r=1}}^2)^{1/(\gamma-1)} = B \,, \label{v1} \ee
and continuity implies that
\be v_{r=1} \left[ 1+\frac {5\gamma+1}{\gamma-1}
{v_{r=1}}^2\right]^{-(3\gamma+1)/(5\gamma+1)}=A \,. \label{cont} \ee
Equation (\ref{inside}) implicitly contains the dependence of $v$
with $r$ in the inner region of the collective wind. Its left hand
side is an ever increasing function. Thus, for the equality to be
fulfilled for all values of radius (0$<r<$1), the right hand side
of the Equation must reach its maximum value at $r$=1. Deriving
the right hand side of Equation ({\ref{inside}), one can find the
velocity which makes it to be maximum
\be v_{\rm max}= \left( \frac{\gamma-1}{\gamma+1 }\right)^{1/2} \,.
\label{max} \ee
Since $v$ grows in the inner region, the maximum velocity is
reached at $r=1$, and from Equation (\ref{v1}),
\be B= \left( \frac{\gamma-1}{\gamma+1 }\right)^{1/2}
\left( \frac 2{\gamma+1}\right)^{1/(\gamma-1)} .\ee
Continuity (Equation \ref{cont}) implies that the value of $A$ is
\be A= \left( \frac{
\gamma-1}{\gamma+1 }\right)^{1/2} \left( \frac
{\gamma+1}{6\gamma+2}\right)^{(3\gamma+1)/(5\gamma+1)}. \ee
With the former value of $B$, and from Equation (\ref{pin}), if
\be P_\infty < \frac 1\gamma \left( \frac{ \gamma-1}{\gamma+1
}\right)^{1/2} \left( \frac {\gamma+1}{2}
\right)^{\gamma/(\gamma-1)} \frac{\dot M_{\rm assoc} V_w}{4\pi
R_c^2} \,, \ee
the subsonic solution is not attainable (continuity of the
velocity flow is impossible) and the supersonic branch is the only
physically viable. In this regime, the flow leaves the boundary of
the cluster $R_c$ at the local sound speed $v_{\rm max}$ (equal to
1/2 for $\gamma=5/3$) and is accelerated until $v=1$ for $r
\rightarrow \infty$.

Fig.  \ref{cantomodel} shows four examples of the supersonic flow
(velocity and particle density) for a group of stars generating
different values of $\dot M_{\rm assoc}$, $V_w$, and $R_c$, as
given in Table 1. The total mass contained up to 10 $R_c$ is also
included in the Table. A typical configuration of a group of tens
of stars (see Appendix) may generate a wind in expansion with a
velocity of the order of 1000 km s$^{-1}$ and a mass between
tenths and a few solar masses within a few pc (tens of $R_c$). We
consider below hadronic interactions with this matter.

\begin{table}[t]
\begin{center}
\caption{Examples of configurations of collective stellar winds.
The mass is that contained within 10 $R_c$. $n_0$ is
the central density.} \vspace{0.2cm}
\begin{tabular}{lccccc}
\hline
 Model    & $\dot M_{\rm assoc}$  & $V_w$          & $R_c$ & $n_0$& Wind mass\\
          & [M$_\odot$ yr$^{-1}$] &  [km s$^{-1}$] & pc    &  cm$^{-3}$    &[M$_\odot$ ]\\
\hline
    A & $10^{-4}$ & 800 & 0.1 & 210.0 & 0.13\\
    B & $10^{-4}$ & 800 & 0.3 & 23.3 & 0.39\\
    C & $5 \times 10^{-5}$ & 1000 & 0.2 & 20.9 & 0.11\\
    D & $2 \times 10^{-4}$ & 1500 & 0.4 & 13.9 & 0.56\\
    E & $2 \times 10^{-4}$ & 2500 & 0.2 & 33.5 & 0.17\\
\hline \hline
\end{tabular}
\end{center}
\label{cantotable}
\end{table}

\begin{figure*}[t]
\centering
\includegraphics[width=4.cm,height=4.cm]{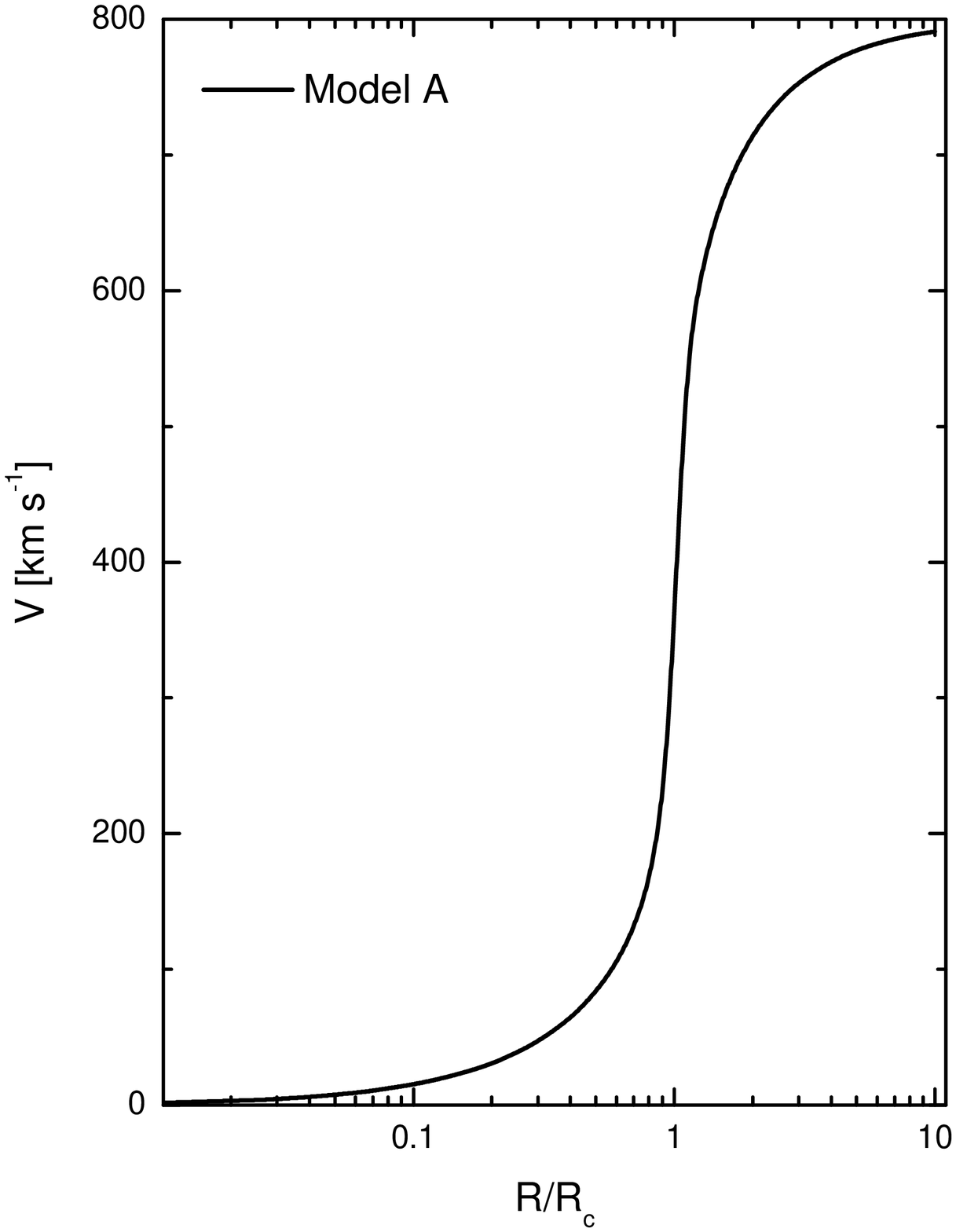}\hspace{.2cm}
\includegraphics[width=4.cm,height=4.cm]{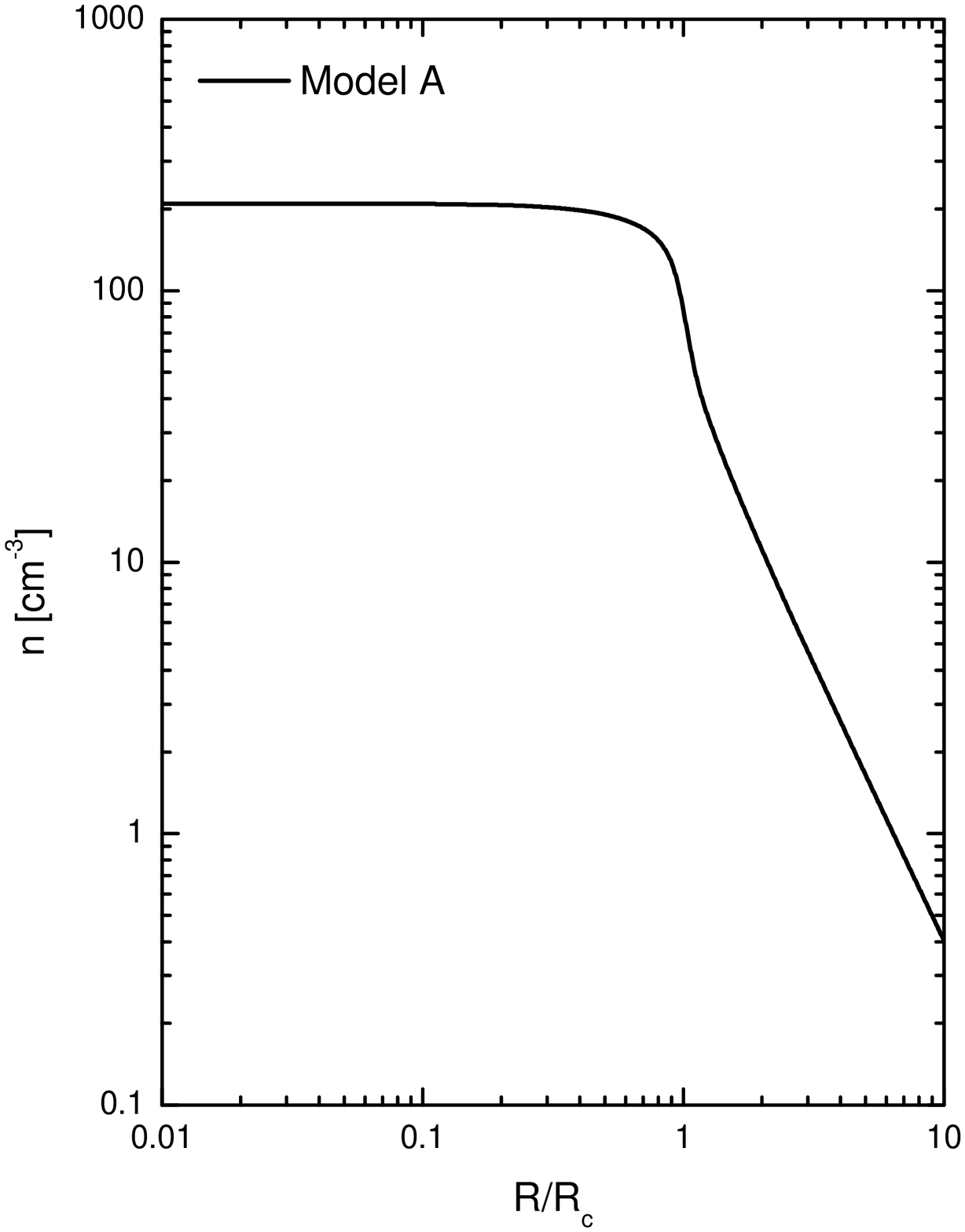}\hspace{.2cm}
\includegraphics[width=4.cm,height=4.cm]{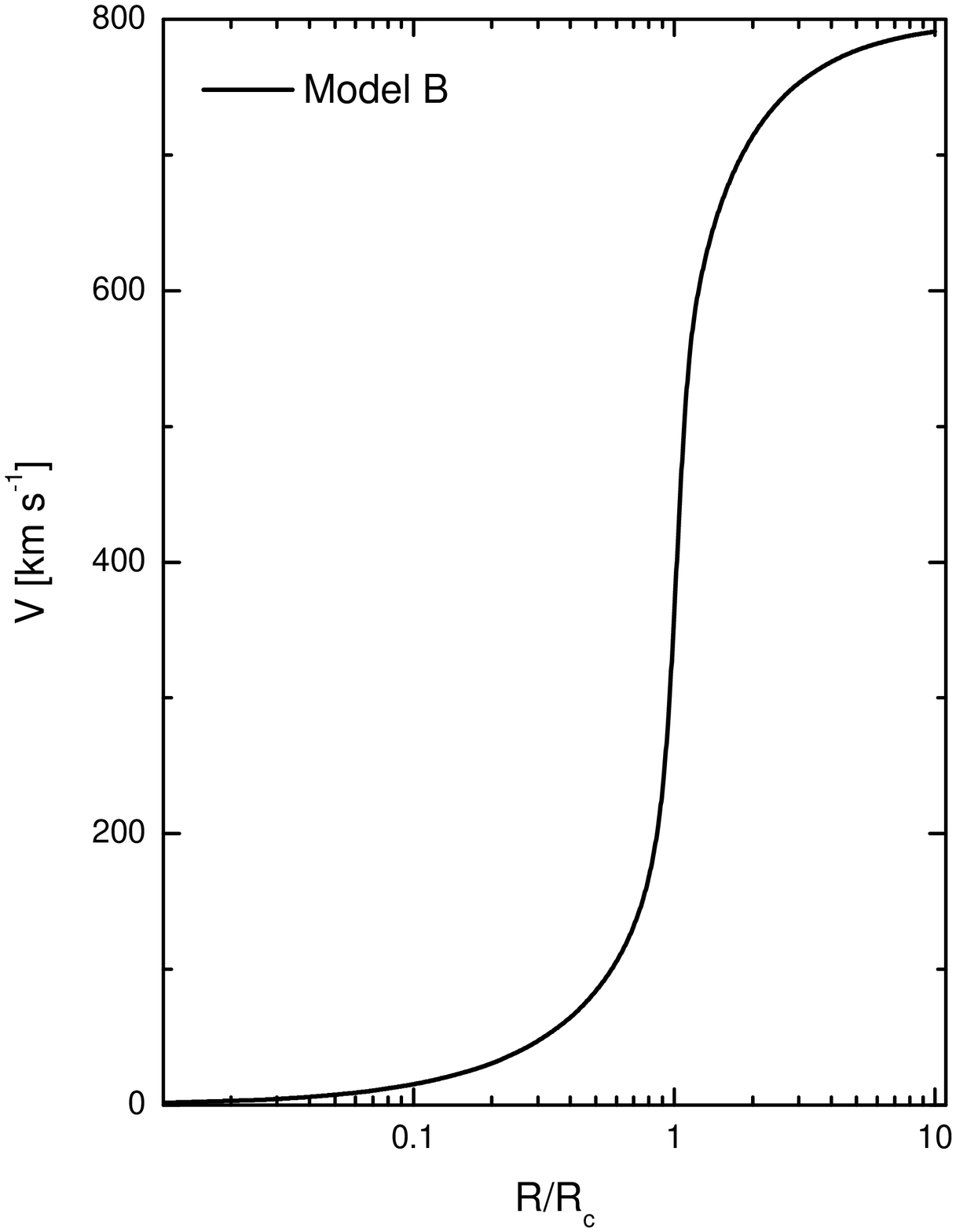}\hspace{.2cm}
\includegraphics[width=4.cm,height=4.cm]{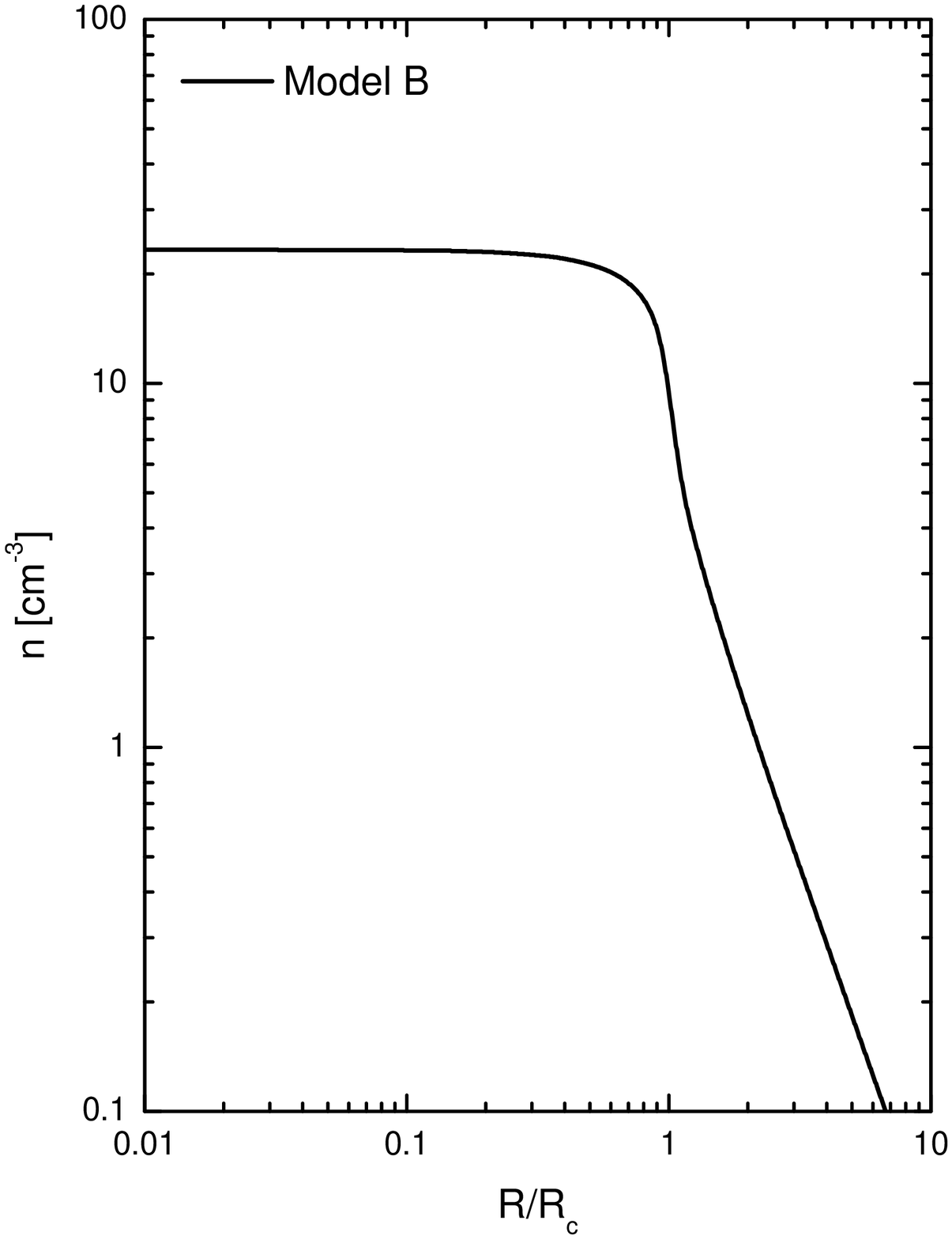}\\
\includegraphics[width=4.cm,height=4.cm]{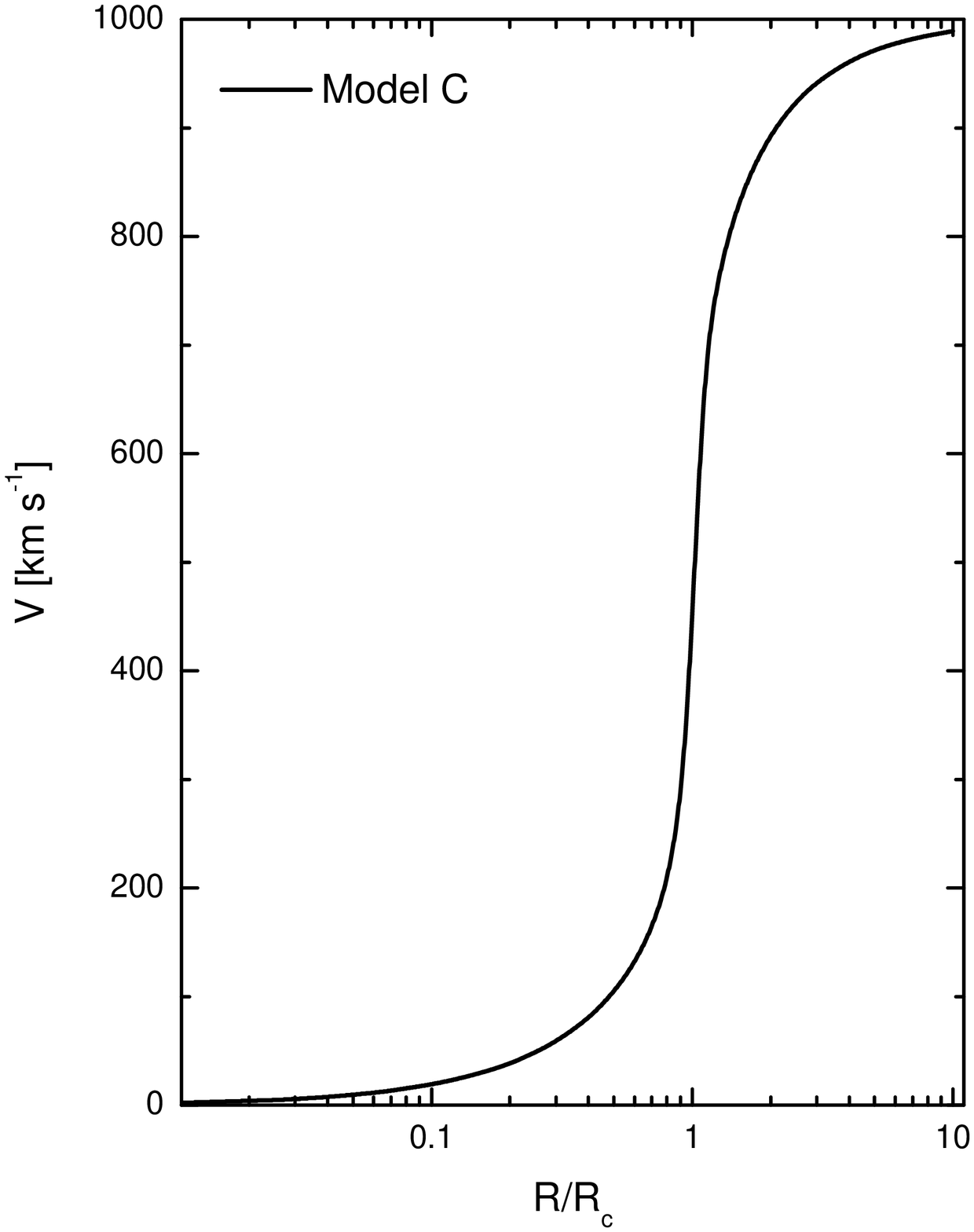}\hspace{.2cm}
\includegraphics[width=4.cm,height=4.cm]{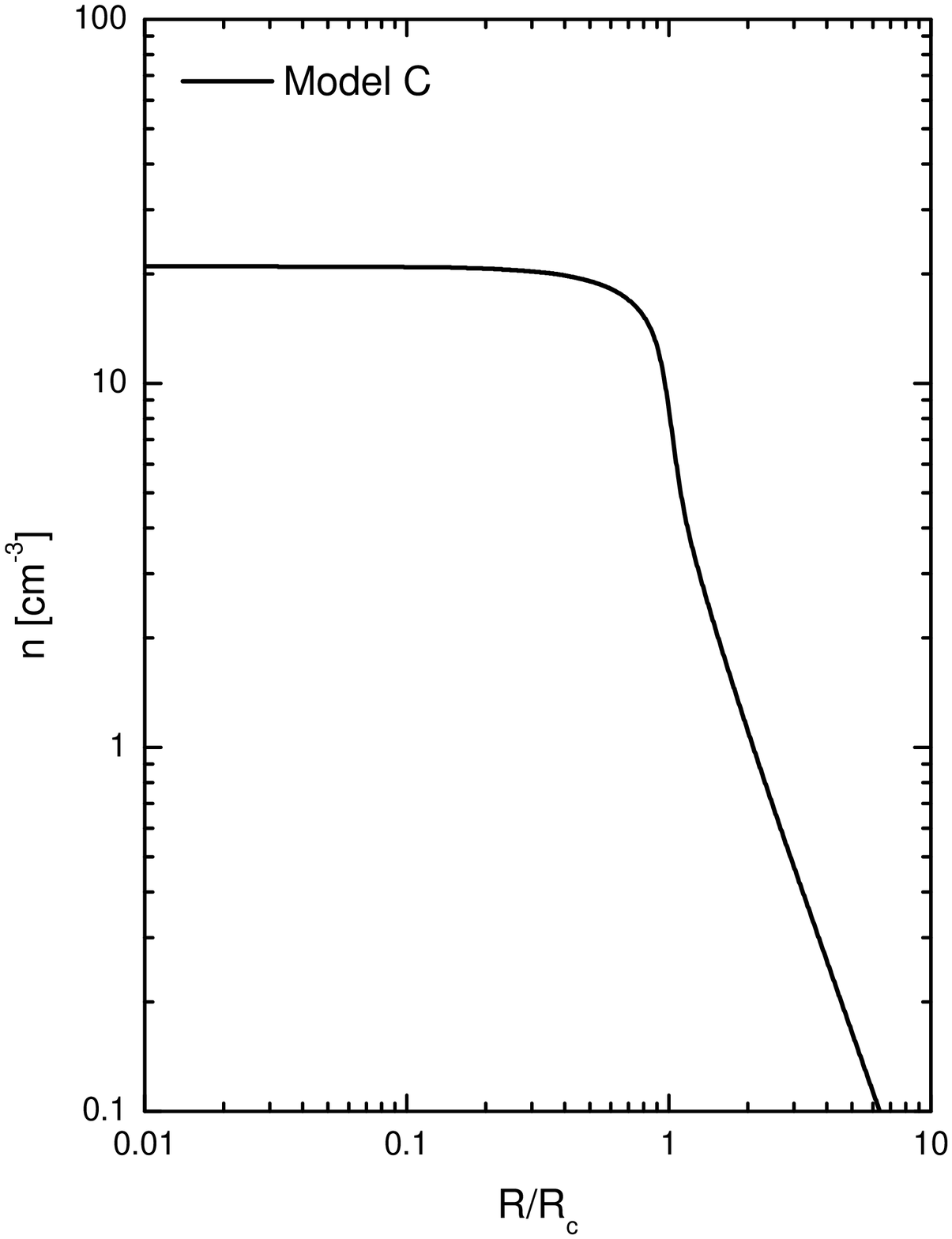}\hspace{.2cm}
\includegraphics[width=4.cm,height=4.cm]{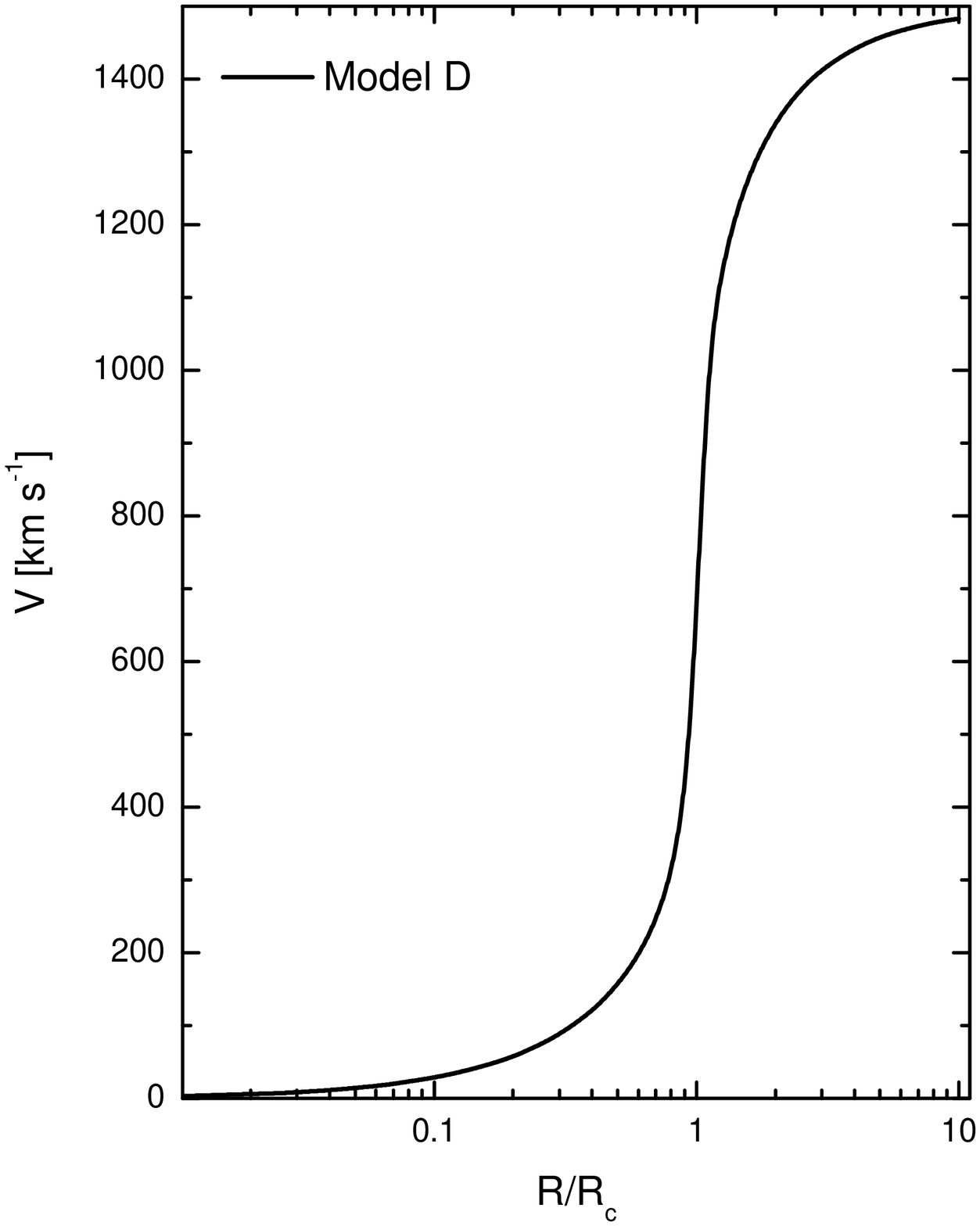}\hspace{.2cm}
\includegraphics[width=4.cm,height=4.cm]{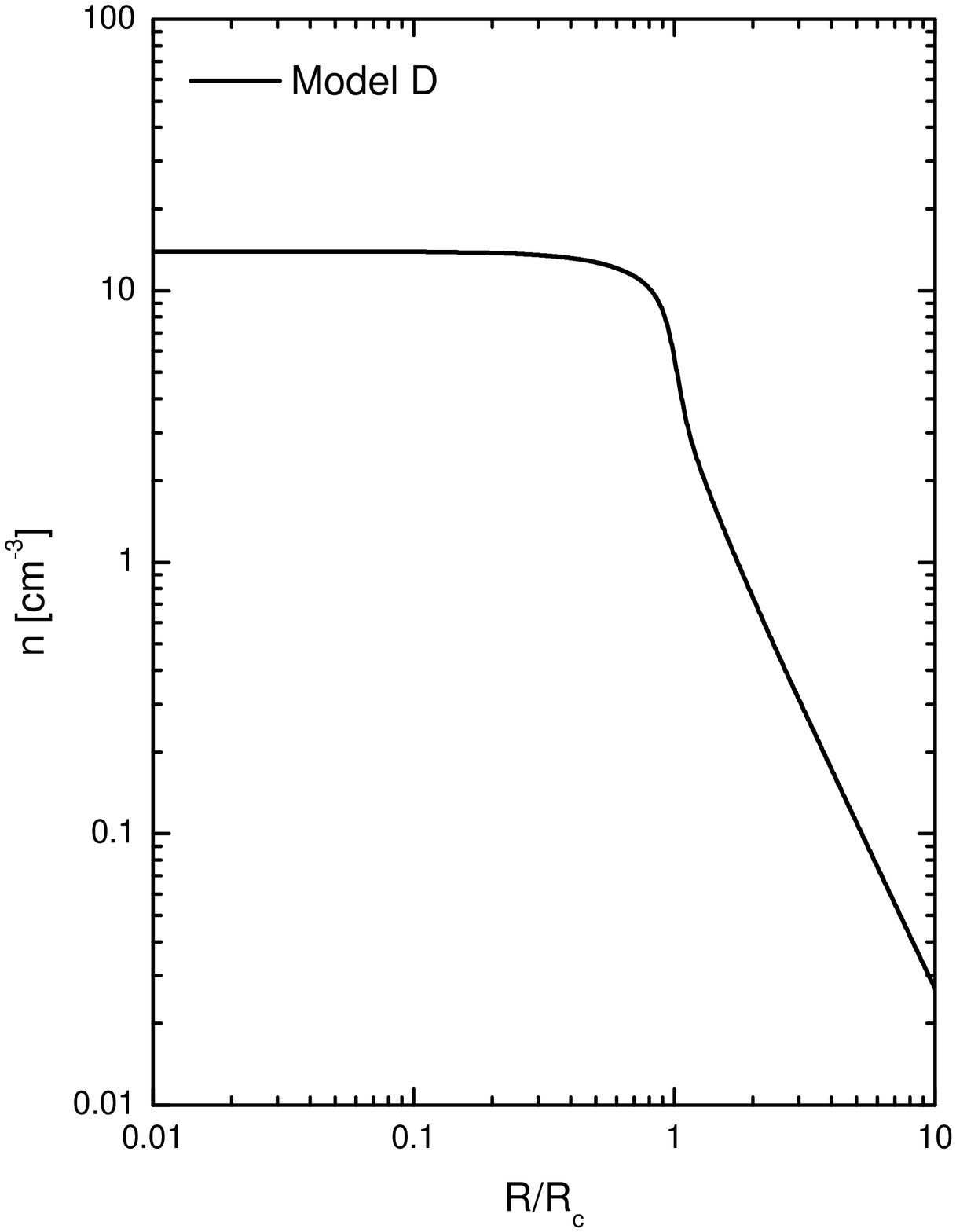}\\
 \caption{Examples of configurations of collective stellar winds. Main parameters
 are as in Table 1.} \label{cantomodel}
\end{figure*}

\section{Modulation and counterparts}

As it happens in the solar system for cosmic rays with less energy
than $\sim 100$ MeV, not all cosmic rays will be able to enter
into the collective wind of several massive stars. The difference
between an {\it inactive target}, as that provided by matter in
the ISM, and an {\it active or expanding target}, as that provided
by matter in a single or a collective stellar wind, is given by
modulation effects.
Although wind modulation has only been studied in detail for the
case of the relatively weak solar wind (e.g. Parker 1958, Jokipii
\& Parker 1970, K\'ota \& Jokipii 1983, Jokipii et al. 1993), and
a proper treatment would have to include a number of different
effects like diffusion, convection, particle drifts, energy
change, and terminal shock barriers, a first approach to determine
whether particles can pervade the wind is to compute the ratio
between the diffusion and convection timescales\be \epsilon
=\frac{t_d}{t_c} = \frac{(3R^2/D)} {(3R/V(R))}.\ee  Here $D$ is
the diffusion coefficient, $R$ is the position in the wind, and
$V$ is the wind velocity. Only particles for which $\epsilon<1$
will be able to overcome convection and enter into the wind region
to produce $\gamma$-rays through hadronic interactions with matter
residing there. A similar approach has also been followed by White
(1985) when computing the synchrotron emission generated by
relativistic particles accelerated in shocks within the wind.
In order to obtain an analytic expression for $\epsilon$ for a
particular star we consider that the diffusion coefficient within
the wind of a particular star is given by (White 1985, V\"olk and
Forman 1982, Torres et al. 2004) \be D \sim \frac 13 \lambda_r c ,
\ee where $\lambda_r$ is the mean-free-path for diffusion in the
radial direction (towards the star). The use of the Bohm
parameterization seems justified, { contrary to what happens in
the solar heliosphere, since we expect that in the innermost
region of a single stellar wind there are many disturbances
(relativistic particles, acoustic waves, radiatively driven waves,
etc.). In the case of a collective wind, the collision of
individual winds of the particular stars forming the association
also produce many disturbances. In any event, a change in the
diffusion coefficient (say a flatter dependence on the energy $E$)
will affect the value of the minimum particle energy that protons
need to enter into the interacting region (see below). We have
proven that unless changes in $E_{\rm min}$ are extreme results
are not significantly affected.}

The mean-free-path for scattering parallel to the magnetic field
($B$) direction is considered to be $\lambda_\| \sim 10 r_g = 10
E/eB$, where $r_g$ is the particle gyro-radius and $E$ its energy.
In the perpendicular direction $\lambda$ is shorter, $\lambda_\bot
\sim r_g$. The mean-free-path in the radial direction is then
given by $\lambda_r={\lambda_\bot}^2 \sin^2 \theta +
{\lambda_\|}^2 \cos^2 \theta = r_g ( 10 \cos^2 \theta + \sin^2
\theta)$, where $\cos^{-2} \theta = 1+(B_\phi/B_r)^2$. Here, the
geometry of the magnetic field for a single star is represented by
the magnetic rotator theory (Weber and Davis 1967; see also White
1985; Lamers and Cassinelli 1999, Ch.~9) \be
\frac{B_\phi}{B_r}=\frac{V_\star}{V_\infty}
\left(1+\frac{R}{R_\star} \right) \label{bp} \ee and \be
\label{br} B_r=B_\star \left(\frac{R_\star}{R}\right)^2 , \ee
where $V_\star$ is the rotational velocity at the surface of the
star, and $B_\star$ the surface magnetic field. Near the star the
magnetic field is approximately radial, while it becomes
tangential far from the star, where $\lambda_r$ is dominated by
diffusion perpendicular to the field lines. This approximation
leads ---when the distance to the star is large compared with that
in which the terminal velocity is reached, what happens at a few
stellar radii--- to values of magnetic field and diffusion
coefficient normally encountered in the ISM.

Using all previous formulae, \ba E^{\rm min}(r) \sim \frac{ 3 e
B_\star V_\infty (r-R_\star)}{ c} \left( \frac {R_\star}{
r}\right)^2 \times \hspace{1.8cm} \nonumber \\  \frac{\left (
1+\left( \frac { V_\star }{V_\infty} \left( 1+ \frac r{R_\star}
\right) \right)^2 \right) ^{3/2}  } {10+ \left( \frac{ V_\star
}{V_\infty} \left( 1+ \frac r{R_\star} \right) \right)^2 } .
\label{ep} \ea Equation (\ref{ep}) defines a minimum energy below
which the particles are convected away from the wind. $E^{\rm
min}(r)$ is an increasing function of $r$, the limiting value of
the previous expression being \ba E^{\rm min}(r \gg R_\star)
&\sim& \frac{ 3 e B_\star V_\infty R_\star}{ c} \left(
\frac{V_\star}{V_\infty} \right) \nonumber \\ & \sim & 4.3 \left(
\frac{B_\star}{10 {\rm G}}\right)  \left(\frac{V_\star}{0.1
V_\infty}\right) \left(\frac{R_\star}{12 {\rm R}_\odot}\right){\rm
TeV } \,.\ea
Therefore, particles that are not convected in the outer regions
are able to diffuse up to its base. Note that $E^{\rm min}(r \gg
R_\star)$ is a linear function of all $R_\star$, $B_\star$ and
$V_\star$, which is typically assumed as $ V_\star \sim 0.1
V_\infty$ (e.g., Lamers and Cassinelli 1999). There is a large
uncertainty in these parameters though, which is safe to consider
about one order of magnitude. The values of the magnetic field in
the surface of O and B stars is a topic of lively discussion.
Despite deep searches, only $5$ stars were found to be magnetic
(with sizeable magnetic fields in the range of $B_\star \sim 100$
G) (e.g.,  Henrichs et al. and references therein), typical
surface magnetic fields of OB stars are then presumably smaller.

In the kind of collective wind we analyzed in Section 2, a first
estimation of the order of magnitude of the energy scale $E^{\rm
min}$ can be obtained as follows. We consider that the collective
wind behaves as that of a single star having a radius equal to
$R_c$, and mass-loss rate equal to that of the whole association,
i.e., $\dot M_{\rm assoc}$. The wind velocity at $R_c$, $V_\star$
is given by Equation (\ref{max}). The order of magnitude of the
{\it surface } magnetic field (i.e., the field at $R=R_c$) is
assumed as the value corresponding to the normal decay of a single
star field located within $R_c$, for which a sensitive assumption
can be obtained using Equations (\ref{bp}) and (\ref{br}), ${\cal
O}(10^{-6}$) G. This results, for the whole association, in
\be [E^{\rm min}(r \gg R_\star)]^{\rm assoc} \sim 0.8 \left(
\frac{B(R_c)}{1 \mu{\rm G}}\right) \left(\frac{R_c}{0.1 {\rm
pc}}\right){\rm TeV } . \label{EminAssoc} \ee
The value of magnetic field is close to that typical of the ISM,
and is bound to be consider as an average (this kind of magnetic
fields magnitude was also used in modelling the unindentified
HEGRA source in Cygnus, see below and Aharonian et al. 2005b). In
particular, if a given star is close to $R_c$ its contribution to
the overall magnetic field near its position will be larger, but
at the same time, its contribution to the opposite region (distant
from it 2 $R_c$) will be negligible. In what follows we consider
hadronic processes up to 10 -- 20 $R_c$, so that a value of
magnetic field typical of ISM values is expected. We shall
consider two realizations of $[E^{\rm min}(r \gg R_\star)]^{\rm
assoc}$, 100 GeV and 1 TeV.

\section{$\gamma$-rays and secondary electrons from
       a cosmic ray spectrum with a low energy cutoff }

The pion produced $\gamma$-ray emissivity is obtained from the
neutral pion emissivity 
as { described in detail in the appendix of
Domingo-Santamar\'{\i}a \& Torres 2005. }
%
%

\subsection{The normalization of the cosmic ray spectrum}

For normalization purposes, we use the expression of the energy
density that is contained in cosmic rays, $\omega_{\rm CR} =
\int_E N(E)\, E\, dE$
%
%
and compare it to the energy contained by cosmic rays in the Earth
environment, $ \omega_{\rm CR,\oplus}(E)=\int_E N_{\oplus}(E)\,
E\, dE $, where $N_{p\,\oplus}$ is the  local cosmic ray
distribution obtained from the measured cosmic ray flux. We assume
the Earth-like spectrum to be $J_\oplus(E)=2.2 E_{\rm GeV}^{-2.75}
{\rm cm^{-2} s^{-1} sr^{-1} GeV^{-1}} $ (e.g. Aharonian et al.
2001, Dermer 1986), so that $ \omega_{\rm CR,\oplus} (E>1 {\rm
GeV}) \sim 1.5 \, {\rm eV cm^{-3}}$. This implicitly defines an
enhancement factor, $\varsigma$, as a function of energy \be
\varsigma(E)=\frac{\int_E N(E) \, E \, dE }{\omega_{\rm
CR,\oplus}(E)} . \label{K} \ee
We assume that $N(E)$ is a power law of the form $N(E) = K_p
E^{-\alpha}$. Values of enhancement $\gg 100$ at all energies are
typical of star forming environments (see, e.g., Bykov \&
Fleishman 1992a,b; Bykov 2001; Suchkov et al. 1993, V\"olk et al.
1996, Torres et al. 2003, Torres 2004; Domingo-Santamar\'{\i}a \&
Torres 2005) and they would ultimately depend on the spectral
slope of the cosmic ray spectrum and on the power of the
accelerator. For a fixed slope, harder than that found in the
Earth environment, the larger the energy, the larger the
enhancement, due to the steep decline ($\propto E^{-2.75}$) of the
local cosmic ray spectrum. In what follows, as an example, we
consider enhancements of the full cosmic ray spectrum (for
energies above 1 GeV) of 1000. With such fixed $\varsigma$, the
normalization of the cosmic ray spectrum, $K_p$, can be obtained
from Equation (\ref{K}) for every value of slope. Note that $K_p
\propto \varsigma$, and thus the flux and $\gamma$-ray luminosity,
$F_\gamma$ and $L_\gamma$, are linearly proportional to the cosmic
ray enhancement.

\subsection{Emissivities}

\begin{figure}[t]
\centering
\includegraphics[width=5cm,height=6cm]{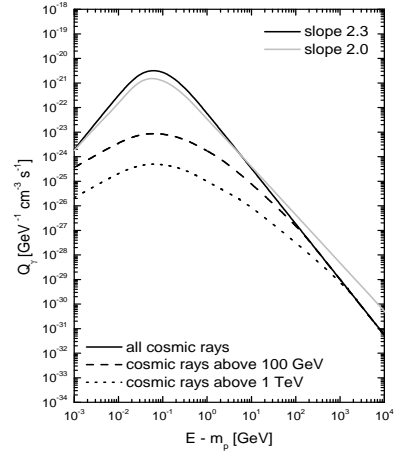}
\caption{Contribution of cosmic rays of different energies to the
hadronic $\gamma$-ray emissivity. The medium density is normalized
to 1 cm$^{-3}$ and the cosmic ray spectrum is proportional to
$E^{-2.3}$ (black) and $E^{-2.0}$ (grey), with an enhancement of a
thousand when compared with the Earth-like one above 1 GeV. The
normalization of each spectrum (of each slope) is chosen to
respect the value of enhancement. In the case of the harder
spectrum of $\alpha=2.0$, we show only the results for the whole
cosmic ray spectrum, but a similar decrease in emissivity to that
of $\alpha=2.3$ can be observed if lower energy cutoffs are
imposed.} \label{f1}
\end{figure}

We now proceed to the computation of the $\gamma$-ray emissivity
produced by a power law spectrum, with a cosmic ray enhancement of
1000 above 1 GeV. As an example, we have assumed $\alpha=2$ and
$2.3$. Fig.  \ref{f1} shows the results for the whole spectrum of
cosmic rays (all cosmic rays, i.e., energies above 1 GeV) and for
cases when the spectrum has a low energy cutoff (100 GeV and 1
TeV).
%
%
For the production of secondary electrons, knock-on and pion
processes need to be taken into account.
{ Details of the computation can be found in Torres 2004. }
%
%
%
%
%
%
%
Numerical results for the knock on emissivity are shown in Fig.
\ref{f3-f4-f5} { left panel} for the same spectra we have used
before. If the cosmic ray spectrum is modulated, the low energy
yield of knock-on electrons gets dramatically reduced. Only when
the electron energies ($E_e$) are sufficiently large so that the
minimum proton energy required to generate them ($E_p^{min}$) is
larger than the modulation threshold, the emissivities obtained
with and without modulation converge.
\begin{figure*}[t]
\centering
\includegraphics[width=5cm,height=6cm]{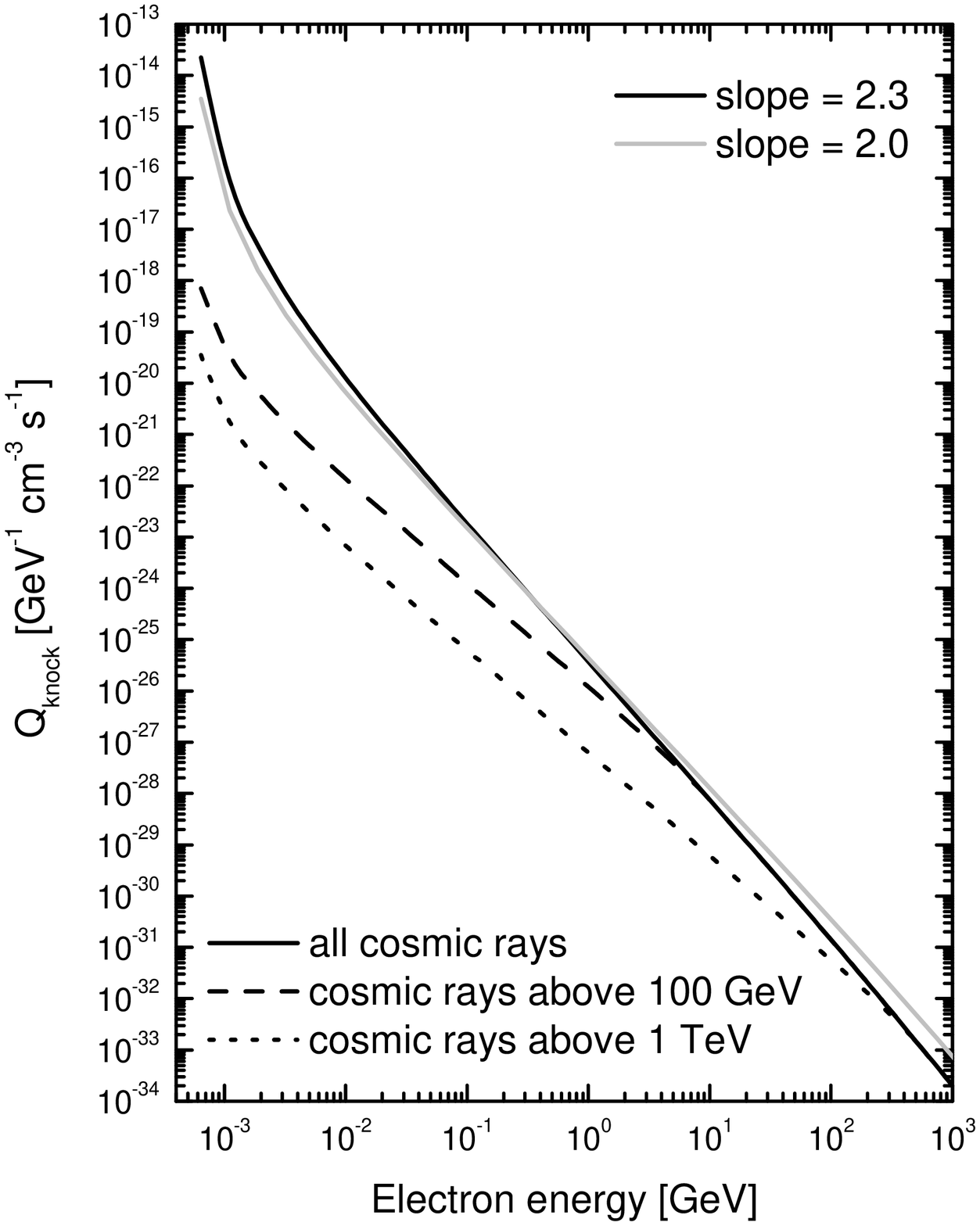}
\includegraphics[width=5cm,height=6cm]{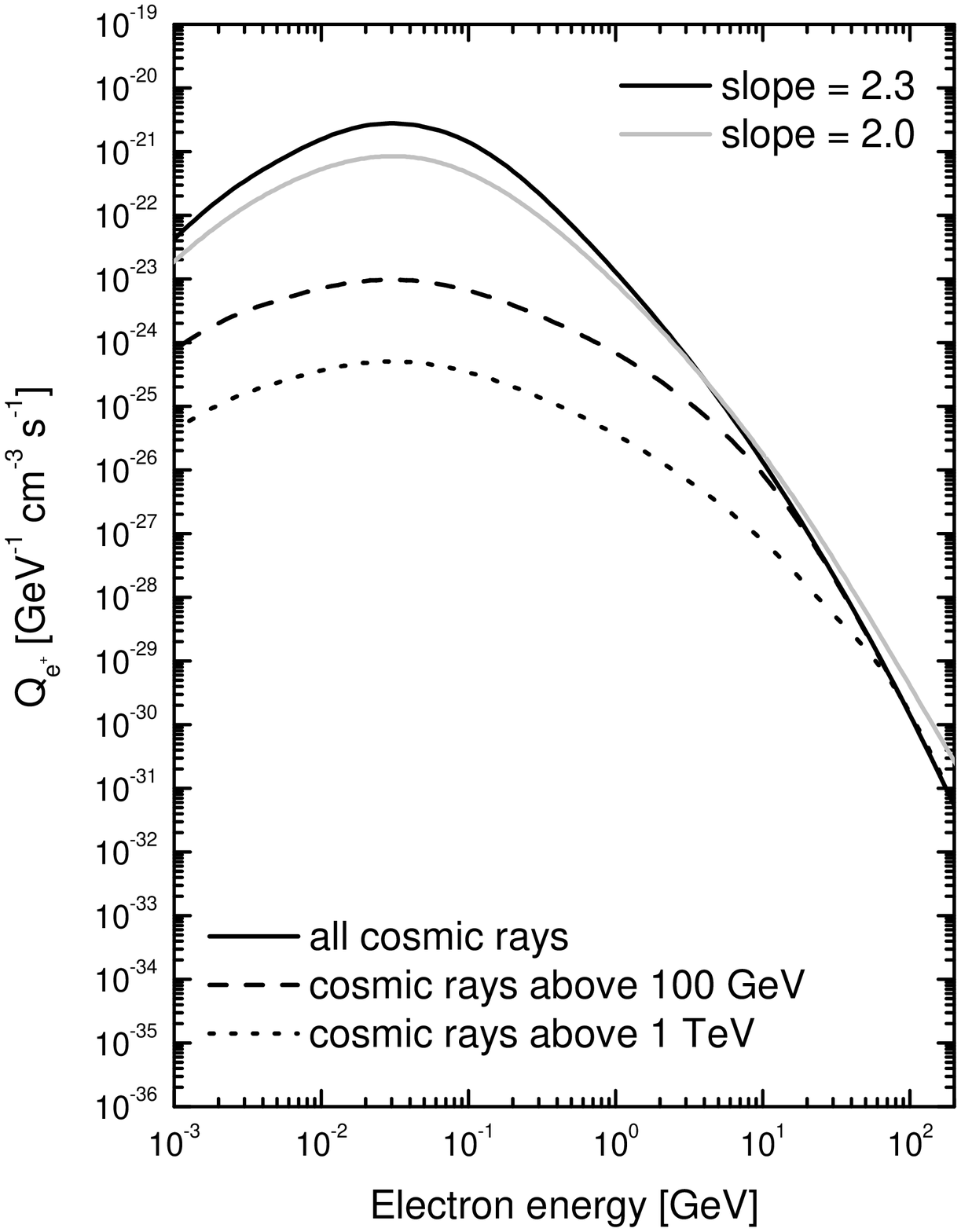}
\includegraphics[width=5cm,height=6cm]{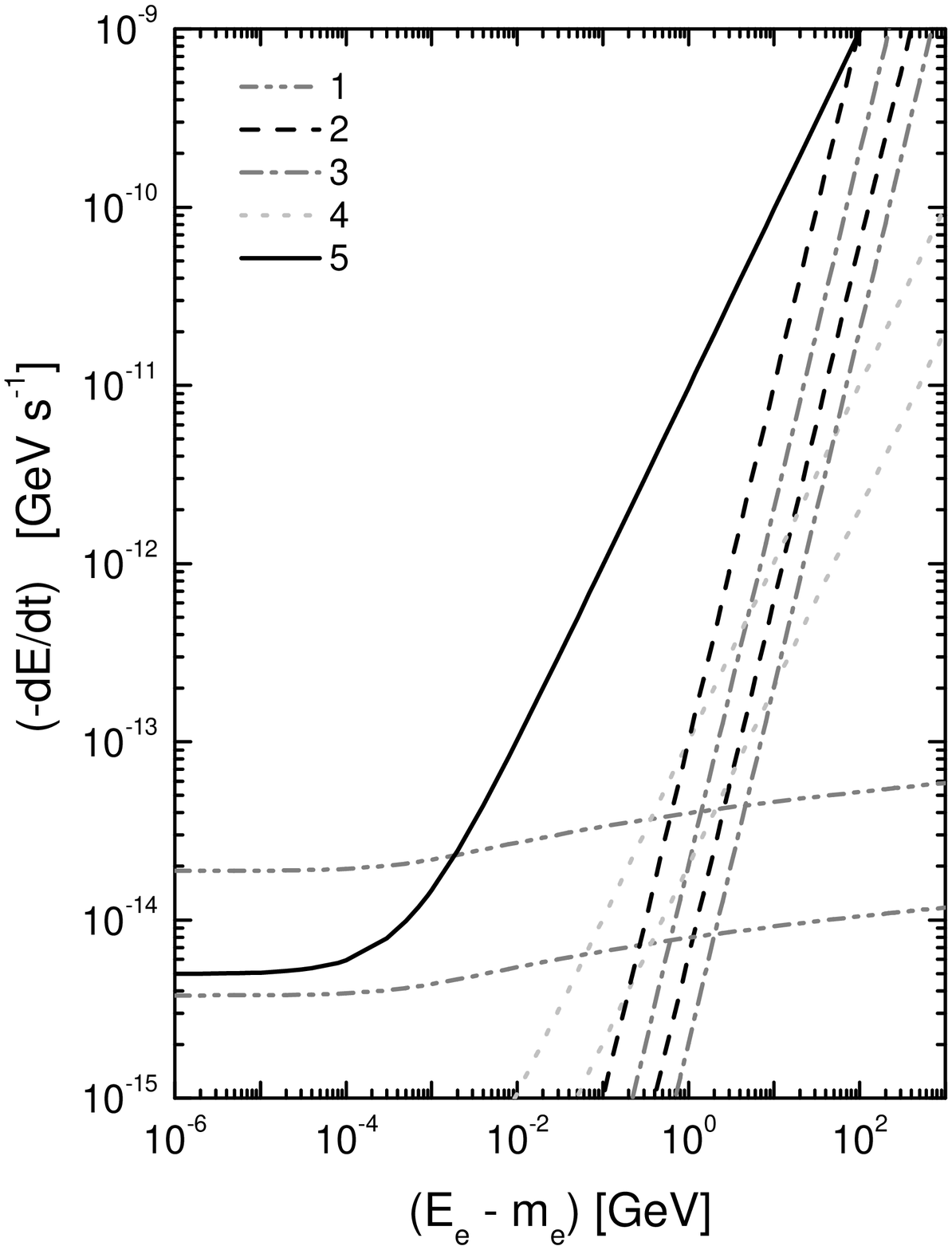}
\caption{The effect of a modulated cosmic ray spectrum (the same
as in Fig.  \ref{f1} with $\alpha=2.3$ and 2.0) over the electron
knock-on emissivity { ({\it left}) } and the positron emissivity {
({\it middle}) } of a medium with density $n=1$ cm$^{-3}$. {
Right: } Different losses for assumed parameters: Curves 1
correspond to ionization losses for $n=100$ and 20 cm$^{-3}$.
Curves 2 correspond to synchrotron losses for $B=50$ and 200
$\mu$G { (see text)}. Curves 3 correspond to inverse Compton
losses for a photon energy density of 20 and 100 eV cm$^{-3}$.
Curves 4 correspond to Bremsstrahlung losses for $n=100$ and 20
cm$^{-3}$. Finally, curve 5 correspond to adiabatic losses having
a ratio $V$(km s$^{-1}$)/$R$(pc)=300 (e.g., a wind velocity of
1500 km s$^{-1}$ and a size relevant for the escape of the
electrons of 5 pc).} \label{f3-f4-f5}
\end{figure*}
For the charged pion emissivity, as an example, we present
numerical results for the case of
positrons in Fig.  
\ref{f3-f4-f5} { middle panel}. As in the case of the neutral pion
photon emissivity, a modulated spectrum would produce a much
smaller electron and positron emissivities at relatively low
energies. This difference can reach several orders of magnitude
when comparing with the spectrum obtained when all cosmic rays get
to interact.

\subsection{Electron energy losses and distribution}


Having the emissivities of secondary electrons we calculate the
electron distribution solving the diffusion-loss equation. This is
$
 \frac{N(E)}{\tau(E)} - \frac{d}{dE} \left[ b(E) N(E) \right] =
Q(E), \label{DL} $
where $Q(E)$ represents all the source terms appropriate to the
production of electrons and positrons with energy $E$, $\tau(E)$
stands for the confinement timescale, $N(E)$ is the distribution
of secondary particles with energies in the range $E$ and $E+dE$
per unit volume, and $b(E)=-\left( {dE}/{dt} \right)$ is the rate
of energy loss. The energy losses considered are those produced by
ionization, inverse Compton scattering, bremsstrahlung,
synchrotron radiation, and the expansion of the medium (for a
compilation of the relevant formulae see the appendix in Torres
2004). A number of uncertain parameters enters into the
computation of these losses. Most notably, these parameters are
the medium density $n$ affecting bremsstrahlung and ionization
losses, the magnetic field $B$ affecting synchrotron losses, the
photon target field affecting inverse Compton losses, and the
velocity of the expanding medium and the size relevant for escape,
affecting the expansion losses.

Fig.  \ref{f3-f4-f5} { right panel} shows the rate of energy
losses for a range of parameters. We show results both $n=100$ and
20 cm$^{-3}$ and we allow magnetic fields  to reach up to 200
$\mu$G in the collective wind region.  { The latter is made in
order to fictitiously enhance --on purpose-- the synchrotron
losses and to better shown the dominance of the other losses
mechanisms over them}. Inverse Compton losses are computed using
two different normalization for the energy density when the photon
target is considered to be a blackbody distribution with $T_{\rm
eff}$=50000 K. The expansion losses have the form \ba
-\left(\frac{dE}{dt} \right)_{{\rm Adia},e} = \frac {V}{R} \left(
\frac{E}{{\rm GeV}}\right){\rm GeV\;s}^{-1}\hspace{3cm} \nonumber \\
= 3.24\times 10^{-14} \left(\frac{V}{100\, {\rm
km\;s}^{-1}}\right) \left(\frac{100{\rm pc}}{R}\right) \left(
\frac{E}{{\rm GeV}}\right){\rm GeV\;s}^{-1}
 \label{ADIA} \ea where $V$ is the collective wind of the region,
and $R$ its relevant size. In Fig.  \ref{f3-f4-f5} { right panel},
to be conservative, we have chosen a ratio $V$(km
s$^{-1}$)/$R$(pc)=300 (e.g., a --single or collective-- wind
velocity of 1500 km s$^{-1}$ and a size relevant for the escape of
the electrons of 5 pc). Fig.  \ref{f3-f4-f5} { right panel} shows
that the expansion dominates the electron losses $b(E)$, as well
as the confinement timescale $\tau (E)$, throughout a wide range
of energies.
In Fig.  \ref{f6} we show an example of the resulting secondary
electron distribution obtained by numerically solving loss
equation. Only at high energies the effect of the cutoff is
unnoticeable, whereas it greatly affects the production of
secondaries (up to several orders of magnitude) below 10 GeV.

\begin{figure}[t]
\centering
\includegraphics[width=5cm,height=6cm]{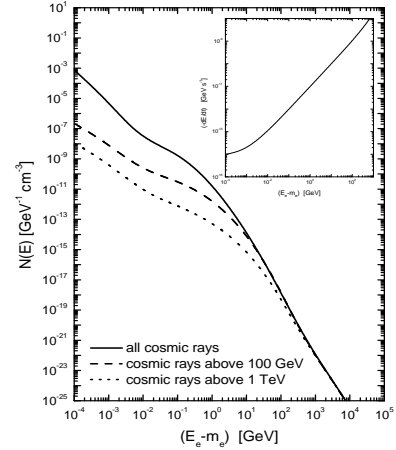}
\caption{Secondary electron distribution obtained by numerically
solving the loss equation. The primary cosmic ray spectrum is the
same as in Fig.  \ref{f1} with $\alpha=2.3$, and results are shown
for different low energy cutoffs. The average density is assumed
as $n=20$ cm$^{-3}$, the magnetic field is assumed as 50 $\mu$G,
and the size relevant for escaping the modulating region is 5 pc.
The inset shows the total energy loss rate $b(E)$.} \label{f6}
\end{figure}


As a benchmark, we note that the radio emission from the electron
population of Fig.  \ref{f6} is well below the upper limits
imposed with VLA at the location of the Cygnus unidentified HEGRA
source (see below for a more detailed discussion, $<200$ mJy at
1.49 GHz, Butt et al. 2003). { Even assuming a rather high
magnetic field of 50 $\mu$G as in Fig. \ref{f6}, what is in the
case of the HEGRA source discarded by X-ray and radio
observations, } we obtain $\sim 50$ mJy at the quoted frequency
for the non-modulated cosmic ray population. 
A smaller magnetic field, as is probably to be found in the outer
regions, or a modulated production of secondaries will diminish
this estimation. We verify too that the limit is respected even
when considering that the primary electron population
(particularly at low energies) is at the same level than the
secondary electron distribution.

\subsection{Total $\gamma$-ray flux from a modulated environment}

To compute $\gamma$-ray fluxes in a concrete example, and
following Section 2, we consider $\sim$2 M$_\odot$ of target mass
being modulated within $\sim$1 pc. The average density is $\sim25$
cm$^{-3}$. This amount of mass is typical of the configurations
studied in Section 2 within the innermost $20\,R_c \sim 2-8$ pc.
To fix numerical values, we consider that the group of stars is at
a Galactic distance of 2 kpc. Using the computations of secondary
electrons and their distribution, the left panel of Fig. \ref{f7}
shows the differential (hadronic and leptonic) $\gamma$-ray flux
when the proton spectrum has a slope of 2.3 and 2.0. In the latter
case, to simplify, we show only the pion decay contribution which
dominates at high energies, as is  produced by the whole cosmic
ray spectrum.

\begin{figure*}[t]
\centering
\includegraphics[width=5cm,height=6cm]{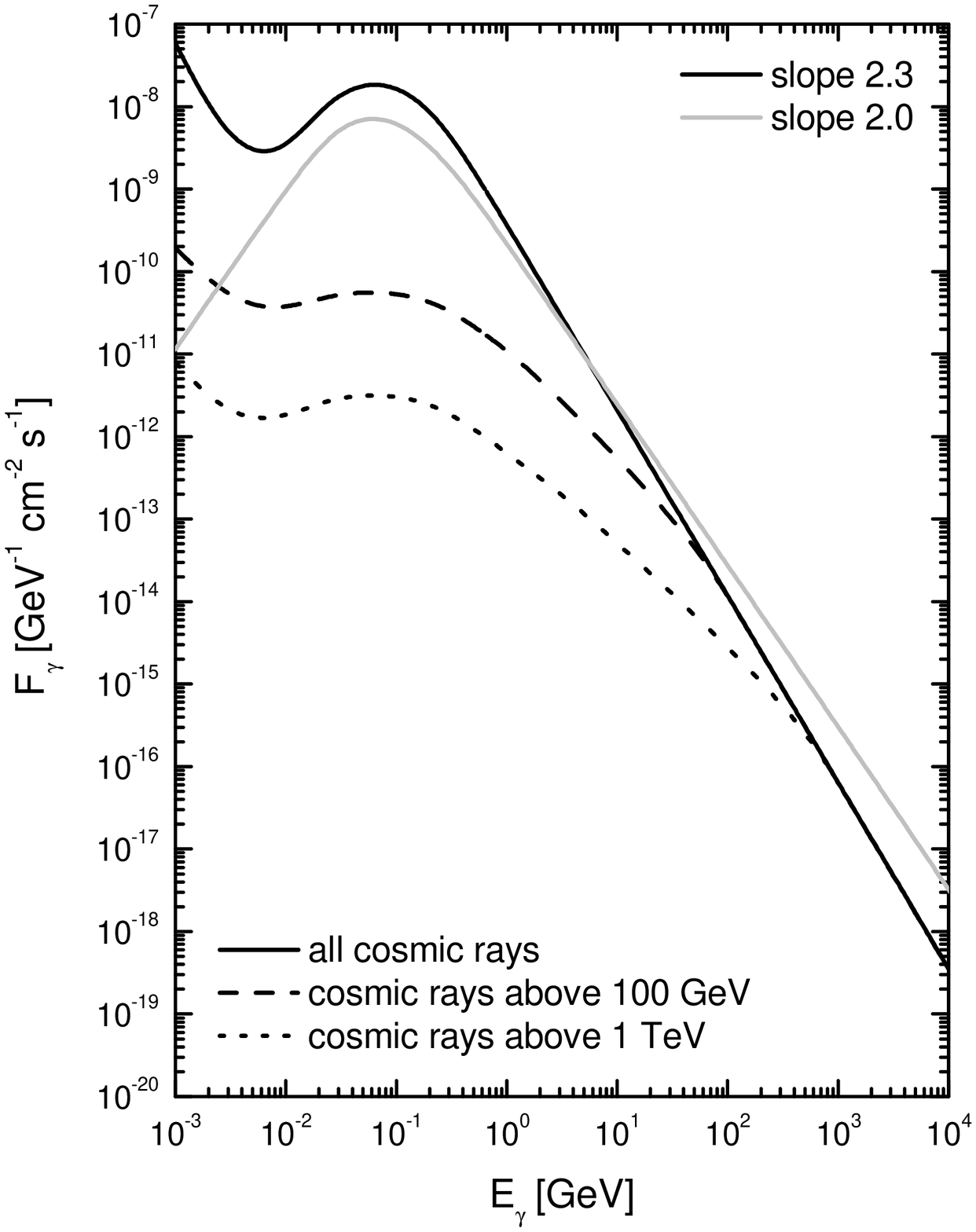}
\includegraphics[width=5cm,height=6cm]{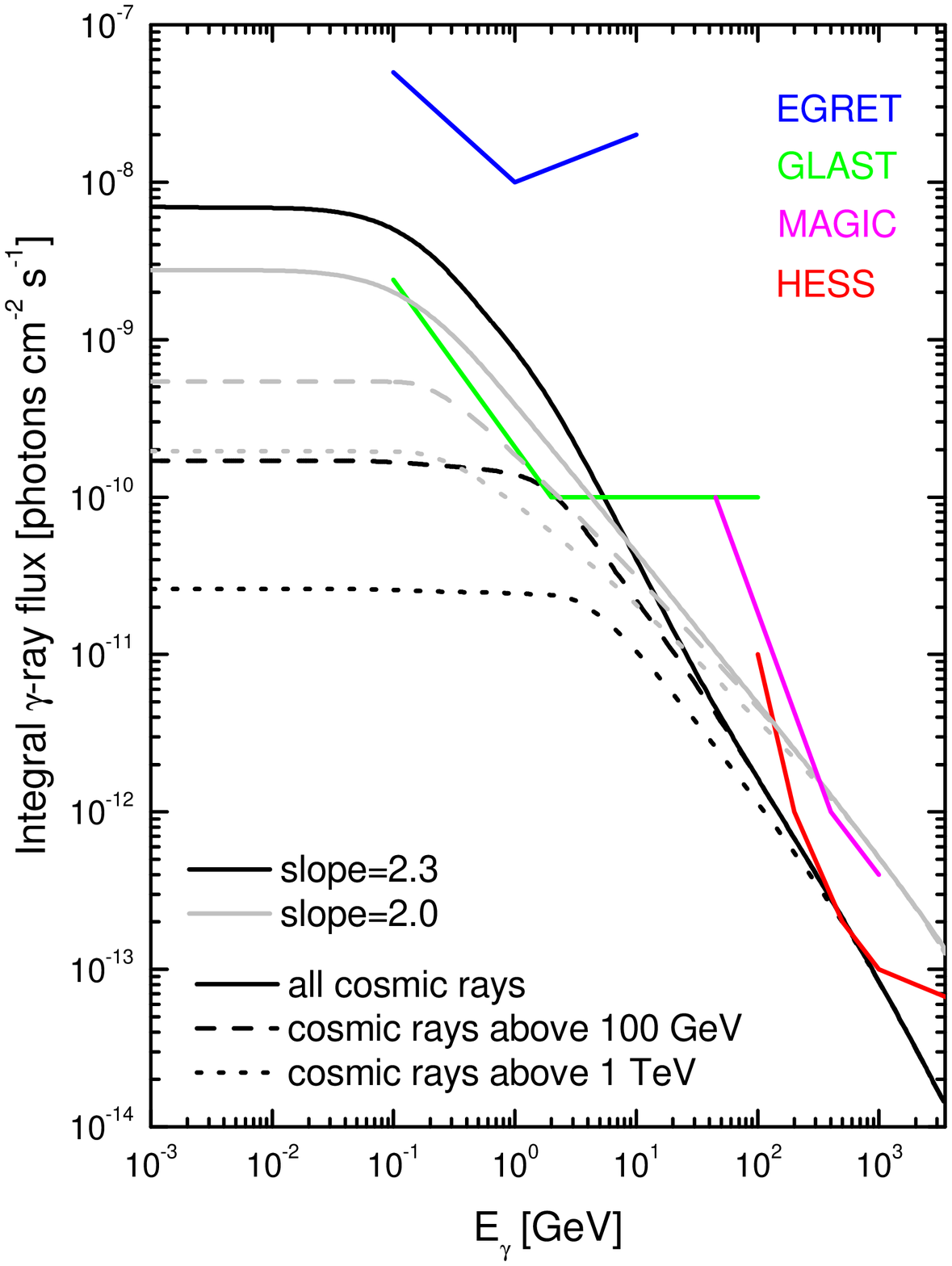}
\includegraphics[width=5cm,height=6cm]{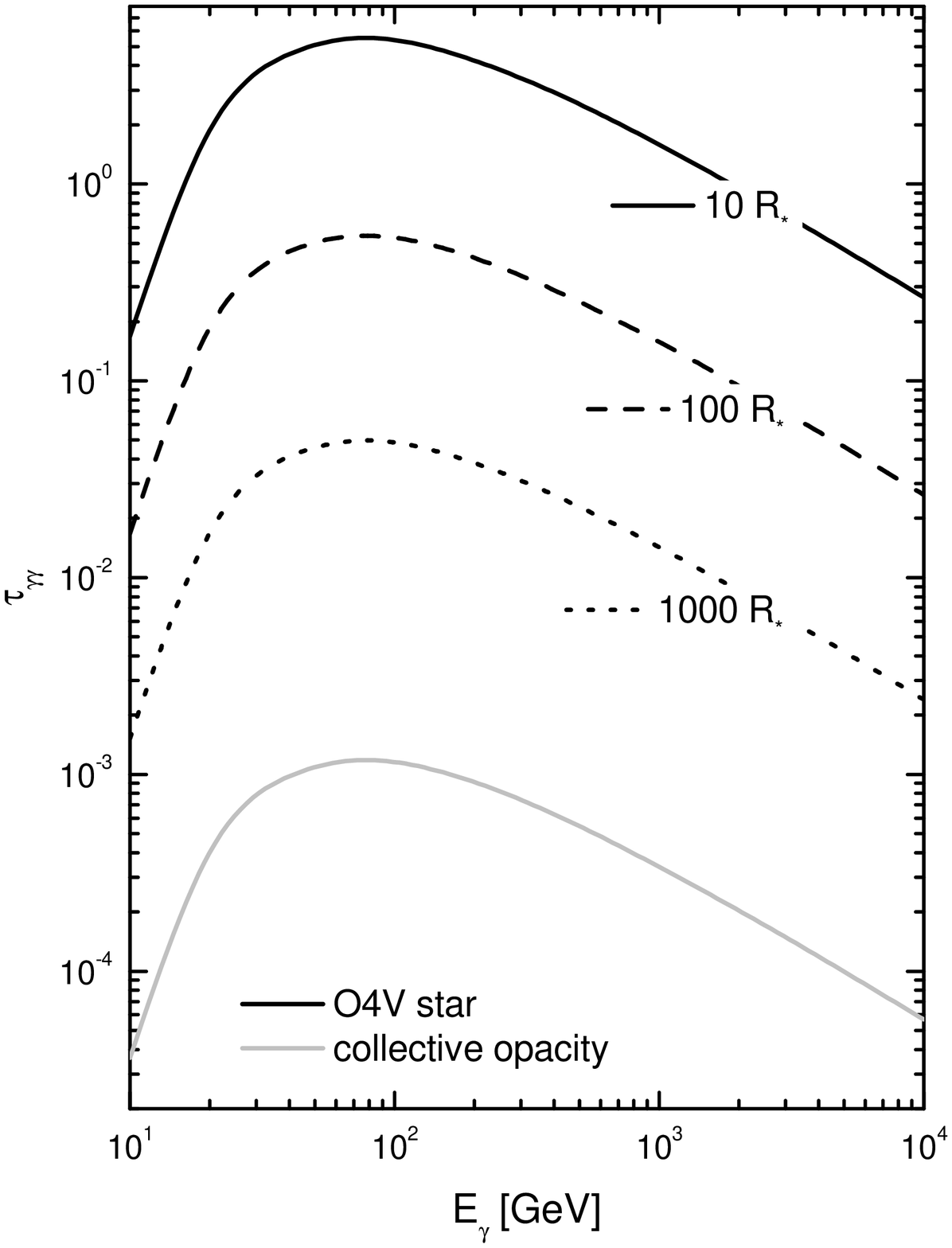}
\caption{Differential (left) and integral (right) fluxes of
$\gamma$-rays emitted in a non-modulated and a modulated
environment. The bump at very low energies in the left panel is
produced because we show leptonic emission coming only from
secondary electrons. Above $\sim 70$ MeV the emission is dominated
by neutral pion decay. Also shown are the EGRET, GLAST, MAGIC, and
HESS sensitivities. Note that a source can be detectable by IACTs,
and not by GLAST, or viceversa, depending on the slope of the
cosmic ray spectrum and degree of modulation. Right: Opacities to
$\gamma\gamma$ pair production in the soft
 photon field of an O4V-star at 10, 100, and 1000 $R_\star$, and in the collective photon field of
 an association with 30 stars distributed uniformly over a sphere
 of 0.5 pc. The closest star to the creation point is assumed to be
 at 0.16 pc, and the rest are placed following the average stellar density
 as follows: 1 additional star within
 0.1, 2 within 0.25, 4 within 0.32, 8 within 0.40 and 14
 within 0.5 pc.} \label{f7}
\end{figure*}

The differential photon flux is given by $ F_\gamma(E_\gamma) =[ {
V}/{4\pi D^2}] {Q_\gamma(E_\gamma)} = [ { M}/{m_p\,4\pi
D^2}][Q_\gamma(E_\gamma)/n] ,$  where $V$ and $D$ are the volume
and distance to the source, and $M$ the target mass. In those
examples where the volume, distance, and/or the medium density are
such that the differential flux and the integral flux obtained
from it above 100 MeV with the full cosmic ray spectrum  is larger
than instrumental sensitivity, a modulated spectrum with a 100 GeV
or a 1 TeV energy threshold might not produce a detectable source
in this energy range. However, the flux will be essentially
unaffected at higher energy (see the high energy end of Fig. s
\ref{f1}, \ref{f3-f4-f5}, and \ref{f6}). The left panel of Fig.
\ref{f7} shows that wind modulation can imply that a source may be
detectable for the ground-based Cerenkov telescopes without being
even close to be detected by instruments in the 100 MeV -- 10 GeV
regime (like the late EGRET or the forthcoming GLAST). To have
further insight on this, the right panel of Fig.  \ref{f7}
presents the integral flux of $\gamma$-rays as a function of
energy, together with the sensitivity of ground-based and
space-based $\gamma$-ray telescopes. The sensitivity curves shown
are for point-like sources, it is expected that extended emission
would require about a factor of 2 more flux to reach the same
level of detectability. Table 3 summarizes these results. From
Table 3 and Fig.  \ref{f7} it is possible to see that there are
plenty of different scenarios (possibly relevant parameters are
distance, enhancement, degree of modulation of the cosmic ray
spectrum, and slope) for which sources that shine enough for
detection at the GLAST domain, may not do so in the IACTs energy
range, and viceversa.

\begin{table}
\begin{center}
\caption{Examples of results for detection in different telescopes
when the configuration of collective stellar winds generates a
target of about 2 M$_\odot$, it is located at 2 kpc, and it is
bombarded with a cosmic ray spectrum having an spectral slope
$\alpha= 2.3$ and 2.0 enhanced a factor of $10^3$ above 1 GeV. The
full cosmic ray spectrum and different modulated cases, at 100 GeV
and 1 TeV, are shown. Except in the case of EGRET, when
$\alpha=2.3$ sensitivities are  barely above the expected fluxes
(see Fig.  (\ref{f7}).} \vspace{0.2cm}
\begin{tabular}{lccc}
\hline
 Telescope   & All  &              100 GeV &              1 TeV\\
 \hline
 $\alpha=2.3$ & \\
 \hline
    EGRET ($E>100$ MeV)         & $\times$ &  $\times$   &$\times$\\
    GLAST ($E>100$ MeV)         & $\surd$  &  $\times$   &$\times$\\
    MAGIC ($E>50$ GeV)          & $\times$ &  $\times$   &$\times$\\
    HESS/VERITAS ($E>100$ GeV) & $\times$ &  $\times$   &$\times$\\
\hline
$\alpha=2.0$ & \\
\hline
EGRET ($E>100$ MeV)        & $\times$     &   $\times$          &  $\times$        \\
GLAST ($E>100$ MeV)        & $\surd$   &   $\times$         &  $\times$       \\
MAGIC ($E>300$ GeV)         & $\surd$   & $\surd$         &  $\surd$    \\
HESS/VERITAS ($E>200$ GeV) & $\surd$   &  $\surd$            &  $\surd$         \\
\hline \hline
\end{tabular}
\end{center}
\label{resul}
\end{table}

\section{Opacity to $\gamma$-ray escape}

The opacity to pair production of the $\gamma$-rays in the UV
stellar photon field of an association can be computed as (Reimer
2003, Torres et al. 2004) \be \label{opa}
\tau(E_\gamma)=\sum_{i=1}^N \int \int_{R_{c,i}}^\infty
N_i(E_\star) \sigma_{e^-e^+}(E_\star,E_\gamma) dE_\star dr_i , \ee
where $E_\star$ is the energy of the soft photons, $E_\gamma$ is
the energy of the $\gamma$-ray, $R_{c,i}$ is the place where the
photon was created with respect to the position of the star number
$i$, $r_i$ is the distance measured from star number $i$, and
$\sigma_{e^-e^+}(E_\star,E_\gamma)$
is the cross section for $\gamma\gamma$ pair production (Cox 1999,
p.214). Note that the lower limit of the integral on $\epsilon$ in
the expression for the opacity is determined from the condition
that the center of mass energy of the two colliding photons should
be such that $(1-(m c^2)^2/(\epsilon \, E_\gamma))
>0$.
The stellar photon distribution of star number $i$ at a position
$r_i$ from the star is that of a blackbody peaking at the star
effective temperature ($T_{{\rm eff }\, i}$) and diluted by the
distance factor, $N_i(E_\star)=(\pi B(E_\star))/(h E_\star c)
\cdot (R_{\star_i}^2)/(r_i^2) \,$,
%
%
where $h$ is the Planck constant, $R_{\star_i}$ is the star number
$i$ radius, and $B(E_\star) = (2{E_\star}^3)/((hc)^2) \cdot
(e^{E_\star/kT_{{\rm eff }\, i}} -1)^{-1} \,$.
%

The size of the emission region (where the target gas mass is
located) that we have considered as an example in the previous
sections is of the order of 1 pc. The typical radius of a massive
star is 10--20 solar radius $\sim$ 5 $\times 10^{-7}$ pc, so that
the most likely creation sites for photons will be far from
individual stars.
In Fig.  \ref{f7} (right panel) we show the value of
$\tau(E_\gamma)$ for different photon creation sites distant from
a O4V-star 10, 100, and 1000 $R_\odot$, with $R_\star = 12R_\odot$
and $T_{\rm eff} = 47400$ K. Unless a photon is created hovering
onto the star, well within 1000 $R_\star$, $\gamma$-ray opacities
are very low and can be safely neglected. This is still true for
associations in which the number of stars is of some tens.
Consider for instance a group of 30 such stars within a region of
0.5 pc (the central core of an association). The stellar density
is given by Equation (\ref{n}), and the number of stars within a
circle of radius $R$ progresses as ${\cal N}= N (R/R_c)^3$. Fig.
 \ref{f7} (right panel) shows the collective contribution to the opacity
obtained from Equation (\ref{opa}) in this configuration, it is
also very low since the large majority of the photons are produced
far from individual stars. This, however, is not the case if one
considers the collective effect of a much larger association like
Cygnus OB 2, particularly at its central region (Reimer 2003).
Reimer demonstrated that even when a subgroup of stars like the
ones considered here is separated from a super cluster like Cygnus
OB2 by about 10 pc, the influence of the latter produces an
opacity about one order of magnitude larger than that produced by
the local stars. Even in this case, Fig.   \ref{f7} (right panel)
shows that this opacity is not enough to preclude escape from the
region of the local enhancement of stellar density. Such is not
the case at the very center of Cygnus OB 2.


\section{Candidates: Cygnus OB 2 and Westerlund 1}

Extensive studies of Galactic and LMC/SMC star clusters and OB
associations suggest that star formation occurs almost
instantaneously (Massey et al. 1995, Leitherer 1999). Typical age
spreads are about 2 Myr or less. This is short in comparison with
stellar evolutionary timescales, except for the most massive
stars, and so a coeval star formation seems appropriate. In
addition, we are particularly interested in the case in which at
the position of the stars that are being illuminated by cosmic
rays there is no current star formation. If it were, the amount of
gas and molecular material in the ISM would be larger than that
contained in the winds, and would make the latter a subdominant
contribution in the generation of the total TeV flux. In
particular, should collision between cosmic rays and nuclei in the
ISM be dominant, there would be no modulation.
This fact and the opacity to $\gamma$-ray escape that is found in
the center of a very massive cluster points to the best scenario:
a sub-group of stars located in the outskirts of an association,
close to an accelerator region, perhaps a SNR, or the association
superbubble itself. The case of Cygnus OB 2 and the TeV source
detected by HEGRA, TeV J2032+4131 (Aharonian et al. 2002, 2005b),
seems to be a possible realization of this scenario. This TeV
source, for which no counterparts at lower energies are presently
identified (Butt et al. 2003), constant during the three years of
data collection, extended (5.6$\pm 1.7$ arcmin, $\sim 2.7 $ pc at
1.7 kpc), separate in about 10 pc (at 1.7 kpc) from the core of
the association, and coincident with a significant enhancement of
the star number density (see Fig. 1 of Butt et al. 2003) might be
suggestive of the scenario outlined in the previous sections. TeV
J2032+4131 presents an integral flux of about 3\% of that of the
Crab [$F_\gamma(E_\gamma>1$TeV$) = 4.5(\pm 1.3) \times 10^{-13}$
photons cm$^{-2}$ s$^{-1}$], and a $\gamma$-ray spectrum $
F_\gamma(E_\gamma)= B ({ E_\gamma}
  /{{\rm TeV}})^{-\Gamma} $  photons cm$^{-2}$ s$^{-1}$ TeV$^{-1}$,
 where  $ B       =  \,4.7\, (\pm2.1_{\rm stat}
 \pm1.3_{\rm sys}) \times 10^{-13} $ and $
 \Gamma  = 1.9  (\pm0.3_{\rm stat} \pm0.3_{\rm sys}).  $
Within the kind of models studied in this work, it would be
possible to explain, apart from consistency with the flux level,
spectrum, and variability, why there is no detectable TeV source
at the central core of Cygnus OB 2 (large opacity to photon
escape), and why there is no EGRET (and will not be a GLAST
source) or other significant radio or X-ray diffuse emission at
the position of the HEGRA detection (modulation of cosmic rays).

However, it is worth noticing that the only stars quoted by Butt
et al. at the position of TeV J2032+4131 are 10 O and 10 B stars,
and not even all of them are within the contours of the source.
The quoted stars are similar, late O and early B, and together
produce a mass-loss rate of about $1\times 10^{-5}$ M$_\odot$
yr$^{-1}$. This mass-loss rate is low and produces not too dense a
collective wind, according to the simulations of Section 2. In
addition, typical distances between these stars are of the order
of a parsec, so that a central core radius is not well defined.
Thus, if we are not missing significant stars in this region,
something we might very well be judging on newest and deeper
Chandra observations of the region (Butt et al, and Reimer et al.
both in preparation) as well as on the expected star density,  a
better approach to study the possible contribution of stellar
winds to this source is just to add that of individual stars
(Torres et al. 2004). If only the currenlty known stars exist, the
produced flux is too low to produce the source unless a larger
enhancement or a low ISM density are invoked. A simple hadronic
scenario where an enhanced spectrum interacts with more abundant
ISM nuclei can not be discarded at this point, since there is no
need to use modulation to explain why there is no source detected
by EGRET (the predicted flux in this model is below EGRET
sensitivity). However, this model can be tested with GLAST
observations (see the example of Fig. \ref{f7}). Also, MAGIC
observations of this region are crucial. We have yet to wait for
further data to reach a definitive answer.

Kn\"odlseder (2003) has compiled a sample of massive young star
clusters (Cygnus clones) that have been observed in the Galaxy
(see his table 2). Westerlund 1 (Wd 1) is a prominent member. The
usually adopted distance to the latter is $D\sim 1.1 \pm 0.4$ kpc
(Piatti et al. 1998), although Clark et al. (2005) favor a
distance between 2--4 kpc. The known zoo of massive stars, clearly
a lower limit in each category, includes 7 WN, 6 WC, 5 Early
transition stars (like LBV, or sgB[e]) and more than 25 OB (Clark
and Negueruela 2002, Clark et al. 2005).  A luminous blue variable
(W243) apparently undergoing an eruption event was also found
(Clark \& Negueruela 2004). W243's mass-loss rate alone can be as
high as $3-6 \times 10^{-4}$ M$_\odot$ yr$^{-1}$. The stellar
population of Wd 1 appears to be consistent with an age of the
order 4--8 Myr if the cluster is coeval, and with a lower limit
for a total mass of about a few thousand M$_\odot$, most likely to
be around a few $\times 10^5$ M$_\odot$ (Clark \& Negueruela 2002,
Clark et al. 2005). Then, Wd~1 is likely to be one of the most
massive young clusters in the Local Group, and in contrast to
Cygnus OB2, it is much more compact (a radius of about 0.6 pc)
--and thus a better target for pointing instruments.   A few stars
a bit far from the center (so that high opacities are avoided)
subject to illumination by a cosmic ray population with a hard
slope would make Wd 1 a $\gamma$-ray source. The HESS observatory
has covered the position of Wd 1 when doing the galactic plane
scan, although with just a few hours of observation time
(Aharonian et al. 2005). We suggest that pointed observations to
Wd 1 might result in its detection and that a model like the one
presented here can be tested jointly by GLAST and HESS.

{ Finally,  special interest on this model may appear in regards
to the serendipitous discovered HESS J1303-631 TeV $\gamma$-ray
source (Aharonian et al. 2005c), extended and still unidentified,
which is in spatial coincidence with the OB association Cen OB6
(see their Fig. 7), that contain al least 20 known O stars and
even 1 WR star.}

\section{Concluding Remarks}

We have studied collective wind configurations produced by a
number of massive stars, and obtained densities and expansion
velocities of the stellar wind gas that is to be target for
hadronic interactions in several examples. We have computed
secondary particle production, electrons and positrons from
charged pion decay, electrons from knock-on interactions, and
solve the appropriate diffusion-loss equation with ionization,
synchrotron, bremsstrahlung, inverse Compton, and expansion losses
to obtain expected $\gamma$-ray emission from these regions,
including in an approximate way the effect of cosmic ray
modulation. Examples where different stellar configurations can
produce sources for GLAST satellite, and the MAGIC/HESS/VERITAS
telescopes in non-uniform ways, i.e., with or without the
corresponding counterparts were shown. Finally, we have commented
on Cygnus OB 2 and Westerlund 1 as two associations where this
scenario could be tested. In the latter case, we have proposed it
to be targeted by HESS.

\section*{Appendix: Mass loss rates and terminal velocities of individual
stars}


To estimate how many stars, and of what kind, are needed to get
certain averages for the association parameters, we adopt the
theoretical wind model C of Leitherer, Robert \& Drissen (1992),
where it is discussed in more detail. This model, referred to as
Model THEOR in Leitherer et al. (1999), is the standard model used
in the synthesis program {\small STARBURST99} to compute
spectrophotometric and statistics of starburst
regions.\footnote{http://www.stsci.edu/science/starburst99} We
remark that mass-loss rates and terminal velocities are indeed
uncertain (typically about $\sim 30$\%). Varying the mass-loss
rate and terminal velocity to those corresponding to other models
will have an impact in any $\gamma$-ray luminosity computation.
Alternative parameterizations to the ones below for $\dot M_\star
$ and $V_\infty$ can be found, for instance, in the work by Vink.
et al. (2000).

\begin{table}[t]
\begin{center}
\caption{Wind model parameters of WR, O, and B stars. }
\vspace{0.2cm}
\begin{tabular}{lrr}
\hline
 Stellar  & $\log [\dot M$ ] &
$V_\infty$   \\
  type & M$_\odot$ yr$^{-1}$ &
 [km s$^{-1}$]  \\
\hline

WNL & -4.2   & 1650  \\
WNE & -4.5   & 1900  \\
WC6-9 & -4.4 & 1800 \\
WC4-5 & -4.7 & 2800  \\
WO & -5.0    & 3500  \\

\hline

O3 & -5.2   & 3190   \\
O4 & -5.4   &  2950 \\
O4.5 & -5.5 & 2900  \\
O5 & -5.6   & 2875  \\
O5.5 & -5.7 & 1960   \\
O6 & -5.8   & 2570  \\
O6.5 & -5.9 & 2455   \\
O7 &   -6.0 & 2295  \\
O7.5 & -6.2 & 1975  \\
O8 &  -6.3  & 1755  \\
O8.5 & -6.5 & 1970  \\
O9 & -6.7   & 1500  \\
O9.5 & -6.8 & 1500    \\
B0 & -7.0   & 1000  \\
B0.5 & -7.2 &  500  \\

\hline

B1 & -7.7  & 500 \\
B1.5 & -8.2&500 \\
B2 & -8.6  & 500 \\
B3 & -9.5  & 500 \\
B5 & -10.0 & 500 \\
B7 & -10.9 & 500 \\
B8 & -11.4 & 500 \\
B9 &-12.0  &500  \\
\hline \hline
\end{tabular}
\end{center}
\label{WR1}
\end{table}

For WR stars\footnote{The first evolutionary phase of a WR star is
the nitrogen-line stage (WNL), which begins when CNO processed
material is exposed at the stellar surface. The second stage is
the WNE, during which no surface hydrogen is detectable. After the
helium envelope is shed, the WC stage occurs, in which strong
carbon lines can be seen, and finally, WO stars are formed, with
high surface oxygen abundances. Further sub-classifications are
made according to line strength ratios (e.g., Smith \& Maeder
1991). }, mass-loss rates are based on observations by van der
Hucht et al. (1986) and Prinja et al. (1990). Average terminal
velocities have been taken from Prinja (1990). For O and B stars,
we shall use the parameterizations of $\dot M$ (M$_\odot$
yr$^{-1}$) and $V_\infty$ (km s$^{-1}$) in terms of stellar
parameters (Leitehrer et al. 1992, 1999), that is also part of
Model THEOR. A multidimensional fit to $\dot M$ (M$_\odot$
yr$^{-1}$) and $V_\infty$ (km s$^{-1}$) as a function of stellar
luminosity $L$, mass $M$, effective temperature $T_{\rm eff}$, and
metallicity $Z$, gives \ba  \log \dot M ({\rm M}_\odot \, {\rm
yr}^{-1}) &=& -24.06 + 2.45 \log [L({\rm L}_\odot)] \nonumber
\\
&&- 1.10 \log [M({\rm M}_\odot)] +1.31 \log [T_{\rm eff} (K)]
\nonumber
\\
&& + 0.80 \log [Z(Z_\odot)] , \label{OM} \ea \ba V_\infty ({\rm km
\, s^{-1} }) &=& 1.23 - 0.30 \log [L({\rm L}_\odot)] +0.55 \log
[M({\rm M}_\odot)] \nonumber
\\  && +0.64 \log [T_{\rm eff}
(K)] + 0.13 \log [Z(Z_\odot)] . \label{OV}\ea In order to obtain
the mass-loss rates and terminal velocities, we then need to apply
a calibration between the spectral sub-types of the O and B stars
and their stellar parameters. For early B and O stars we adopt the
work by Vacca et al. (1996), see their Tables 5-7. Columns 2 and 3
of Table 1 show the stellar wind parameters resulting from the
previous calibration. Since in general, many of the stars in a
young association will pertain to luminosity class V, we give the
results for this class only, although combining this Section with
the work by Vacca et al. (1996), values for other classes could be
obtained as needed. We adopt the values of masses as derived using
evolutionary codes, since as discussed by Vacca et al. (1996), the
cause of the mass discrepancy between results from the former
codes and spectroscopic analysis seems to be that the models used
in the spectroscopic studies do not properly take into account the
effects of the wind extension, mass outflow, and velocity fields,
and are thus less reliable.
%
%
%
For B stars of late type, we complement the calibration by Vacca
et al. (1996) with that of Humphreys \& Mc Elroy (1984), see their
Table 2, and Schmidt-Kaler (1982), which give the bolometric
corrections, ($BC$), and effective temperatures. The total
luminosity is then computed using the equation \be \log
(L/L_\odot)= -0.4(M_v+BC-M_{\rm bol \, \odot}) , \ee where $M_{\rm
bol \, \odot}=4.75$ mag (Vacca et al. 1996). The masses and visual
magnitudes $M_v$ of B stars are obtained following the method
described in Appendix A of Kn\"odlseder (2000).
%
%
%
The star radius of each of the stars can be obtained by the usual
relationship $R_\star^2=L/(4 \pi \sigma T_{\rm eff}^4)$, where
$\sigma=5.67 \times 10^{-5}$ erg s$^{-1}$ cm$^{-2}$ K$^{-4}$.

It is to be noted that this mass-loss parameterization may yield
to an under-estimation for certain stars, and an over-estimation
of the terminal velocities, like for example in the case of
HD\,93129A (Benaglia \& Koribalski 2004). We remark that almost
400 Galactic O stars have recently been compiled in the new
`Galactic O Star Catalogue' (GOS) by Ma\'{\i}z-Apell\'aniz \&
Walborn (2002), but there are only about a dozen stars known with
spectral types O3.5 or earlier, thus limiting the statistical
knowledge for comparison. In any case, this parameterization is
enough to show that a group of several tens of early stars can
easily reach a $\dot M_{\rm assoc} \sim {\cal O}
(10^{-5}-10^{-4}$) M$_{\odot}$ yr$^{-1}$.

\section*{Acknowledgments}

The work of ED-S was done under a FPI grant of the Ministry of
Science and Tecnology of Spain. We acknowledge G. Romero and P.
Benaglia for discussions on this topic. We also acknowledge L.
Anchordoqui, A. Bykov, I. Negueruela, A. Reimer, and O. Reimer for
comments at different stages of this work.

\end{document}